\documentclass[epj]{svjour}
\usepackage{graphics}
\usepackage{amsmath}
\def\eqlab#1{\label{eq:#1}}
\def\eqref#1{Eq.~(\ref{eq:#1})}

\def\figlab#1{\label{fig:#1}}
\def\figref#1{Fig.~(\ref{fig:#1})}

\def\tablab#1{\label{tab:#1}}
\def\tabref#1{Table~(\ref{tab:#1})}

\def\seclab#1{\label{sect:#1}}
\def\secref#1{Section~\ref{sect:#1}}

\begin{document}
\title{Feasibility of constraining the curvature parameter of the symmetry energy using elliptic flow data}
\author{M.D. Cozma\inst{1}% etc
\thanks{\emph{Email address:} dan.cozma@theory.nipne.ro}%
}                     % Do not remove
\institute{Department of Theoretical Physics, IFIN-HH, Reactorului 30, 077125 M\v{a}gurele/Bucharest, Romania}
\date{Received: date / Revised version: date}
\abstract{
A QMD transport model that employs a modified momentum dependent interaction (MDI2) potential, 
supplemented by a phase-space coalescence model fitted to FOPI experimental multiplicities of
free nucleons and light clusters is used to study the density dependence of the symmetry
energy above the saturation point by a comparison with  experimental elliptic flow ratios measured by the
FOPI-LAND and ASYEOS collaborations in $^{197}$Au+$^{197}$Au collisions at 400 MeV/nucleon 
impact energy. A previous calculation using the same model has proven that 
neutron-to-proton and neutron-to-charged particles elliptic flow ratios probe on average different
densities allowing in principle the extraction of both the slope $L$ and curvature $K_{sym}$
parameters of the symmetry energy. To make use of this result a Gogny interaction inspired
potential is modified by the addition of a density-dependent, momentum-independent 
term, while enforcing a close description of the empirical nucleon optical potential, 
allowing independent modifications of $L$ and $K_{sym}$. 
Comparing theoretical predictions with experimental data for neutron-to-proton and 
neutron-to-charged particles elliptic flow ratios the following constraint is extracted:
$L$=85$\pm$22(exp)$\pm$20(th)$\pm$12(sys) MeV and $K_{sym}$=96$\pm$315(exp)$\pm$170(th)$\pm$166(sys) MeV.
Theoretical errors include effects due to uncertainties in the isoscalar part of the equation of state,
value of the isovector neutron-proton effective mass splitting, in-medium effects on the elastic nucleon-nucleon
cross-sections, Pauli blocking algorithm variants and scenario considered for the conservation of
the total energy of the system. Systematical uncertainties are generated by the inability of the
transport model to reproduce experimental light-cluster-to-proton multiplicity ratios. A value 
for $L$ free of systematical theoretical uncertainties can be extracted from the neutron-to-proton elliptic flow ratio alone: 
$L$=84$\pm$30(exp)$\pm$19(th) MeV. It is demonstrated that elliptic flow ratios reach a maximum
sensitivity on the $K_{sym}$ parameter in heavy-ion collisions of about 250 MeV/nucleon impact energy,
allowing a reduction of its experimental component of uncertainty to about 150 MeV.
\PACS{
      {21.65.Mn}{Nuclear matter equations of state}   \and
      {21.65.Cd}{Nuclear matter asymmetric matter} \and
      {25.75.Ld}{Collective flow, relativistic collisions} \and
      {25.70.-z}{Heavy-ion nuclear reactions, low and intermediate energy}
     } % end of PACS codes
} %end of abstract
\maketitle
\section{Introduction}
\seclab{intro}
The density dependence of the isospin dependent part of the equation of state (asy-EoS) of nuclear matter, commonly
known as the symmetry energy (SE), represents one of the remaining open questions in nuclear physics. Its
impact on the structure of rare isotopes, dynamics of heavy-ion collisions, properties of 
astrophysical objects such as neutron stars and explosions of supernovae has motivated a large number of
both theoretical and experimental investigations~\cite{Li:2008gp,Lattimer:2006xb}. 
Experimental studies of isospin diffusion, Pygmy and giant dipole resonances, neutron skin thickness 
and other phenomena has made possible the extraction  of constraints with satisfactory accuracy
for the density dependence of the SE in the vicinity of or below the saturation point
~\cite{Chen:2004si,Chen:2005ti,Famiano:2006rb,Klimkiewicz:2007zz,Trippa:2008gr,Centelles:2008vu,Carbone:2010az,RocaMaza:2011pm,Tsang:2012se,Brown:2013mga,Danielewicz:2013upa}. 
Similarly, recent advances in theoretical many-body simulations of nuclear matter has allowed increasingly
more accurate predictions for the asy-EoS up to densities close to the saturation point~
\cite{Hebeler:2009iv,Gezerlis:2013ipa,Tews:2012fj,Kruger:2013kua,Drischler:2015eba,Drischler:2016djf}. The density
dependence far above saturation, $\rho \ge 2 \rho_0$, has remained up to now extremely uncertain. Its knowledge is
mandatory for a proper understanding of properties of neutron stars such as radius and maximum
allowed mass and may provide the key for a simple solution of the so called hyperon puzzle
~\cite{Vidana:2000ew,SchaffnerBielich:2000yj,Djapo:2008au}.%~\cite{Bombaci:2016xzl}.

It is common practice to perform a Taylor expansion in density of the SE around the saturation point $\rho_0$,
\begin{eqnarray}
\eqlab{taylorexpse}
 S(\rho)=S_0+\frac{L}{3} \frac{\rho-\rho_0}{\rho}+\frac{K_{sym}}{18}
 \Big(\frac{\rho-\rho_0}{\rho}\Big)^2+\cdots \,,
\end{eqnarray}
and express constraints for its density dependence in terms of allowed ranges for the
value of SE at saturation ($S_0$), its slope $L$ and curvature $K_{sym}$ parameters.
The results of the majority of the numerous available studies are compatible with $S_0$=32$\pm$2 MeV
and $L$=60$\pm$20 MeV. In contrast, studies aimed at an explicit determination of the asymmetric
part of the compressibility modulus $K_\tau$, allowing an indirect determination of $K_{sym}$,
lead to biased values for the latter as a consequence of potentially unphysical correlations
between $L$ and $K_{sym}$ induced in $K_\tau$ by the choice of the interaction
~\cite{Shlomo:1993zz,Yoshida:2006nk,Sagawa:2007sp,Piekarewicz:2008nh,Chen:2009wv,Li:2010kfa,Stone:2014wza}.

These correlations can be traced back to the particular expressions for the employed Gogny
and Skyrme type interactions which include minimal density dependent terms introduced as
effective approximations to the 3-body nucleon - nucleon force~\cite{Vautherin:1971aw,Decharge:1979fa}.
Their strength has been fixed by requiring that empirical saturation properties of nuclear matter are
reproduced. Microscopical model calculations using realistic two- and three-nucleon interactions
have however demonstrated the sensitivity of both the value of SE at saturation and the
maximum possible mass of neutron stars to the yet insufficiently constrained part of
the 3-body interaction: intermediate range (3 pion loops) and the spatial and spin structure
of the short-range 3-neutron terms~\cite{Gandolfi:2011xu,Steiner:2011ft}. Empirical information
on the density dependence of the SE above saturation should therefore be extracted using models that 
lift these constraints appearing in early studies as result of using simple parametrizations for the interaction.

Intermediate energy heavy-ion collisions (HIC) provide the unique opportunity to create 
and study in terrestrial laboratories chunks of nuclear matter in the vicinity of
twice saturation density~\cite{Li:2002yda}. Several promising observables have been
identified for this purpose: the ratio of neutron-to-proton yields of squeezed out 
nucleons~\cite{Yong:2007tx}, light cluster emission~\cite{Chen:2003ava}, $\pi^-/\pi^+$ multiplicity
ratio in central collisions~\cite{Li:2004cq}, differential transverse flow~\cite{Li:2002qx} and others.

Elliptic flow ratios (EFR) and differences (EFD) of isospin partners have been shown to be sufficiently sensitive 
to probe the supranormal density dependence of SE~\cite{Russotto:2011hq,Cozma:2011nr} and by making
use of the FOPILAND~\cite{Leifels:1993ir,Lambrecht:1994cp} and, more recently, 
ASYEOS~\cite{Russotto:2016ucm} experimental data constraints for the asy-EoS stiffness 
have been extracted~\cite{Russotto:2011hq,Russotto:2016ucm,Cozma:2013sja,Wang:2014rva}.
All these studies have made use of EoS parametrizations that allow adjustments of
the SE stiffness by modifying the value of a single parameter. Consequently only the slope of the SE
averaged over the probed density region could be extracted. This may in general be different from the
slope at saturation $L$, but well within the quoted uncertainties for the mentioned studies.

Using an upgraded version of the T\"ubingen QMD transport model it has been shown that the
neutron-to-proton elliptic flow ratio (npEFR) and neutron-to-hydrogen elliptic flow ratio(nhEFR) 
probe on average different density regimes, 1.4-1.5$\rho_0$ and 1.0-1.1$\rho_0$ 
respectively~\cite{Russotto:2016ucm}. The case of neutron-to-charged particles elliptic flow ratio (nchEFR)
is similar to that of nhEFR. It may thus be possible to extract constraints for both the slope $L$
and curvature $K_{sym}$ parameters from a comparison of transport model predictions with combined
experimental data for npEFR and nhEFR, or alternatively npEFR and nchEFR.

The present study aims at extracting constraints for both the slope $L$ and curvature $K_{sym}$
from the FOPILAND npEFR and ASYEOS nchEFR experimental data. To that end, the MDI Gogny inspired
effective potential~\cite{Das:2002fr} used to describe the mean-field interaction of nucleons
in the transport model of choice (T\"ubingen QMD) is modified by including an additional
term (MDI2), in a similar fashion to Ref.~\cite{Xu:2014cwa}, allowing independent variations of $L$ and $K_{sym}$. 
The final state spectra of HIC simulations are determined using a minimum spanning tree (MST)
coalescence algorithm. All relevant details of the model are presented in \secref{model}. Predictions
of the model are then compared with published FOPI~\cite{Reisdorf:2010aa,FOPI:2011aa},
FOPI-LAND~\cite{Cozma:2013sja} and ASYEOS~\cite{Russotto:2016ucm} experimental data for
transverse and elliptic flow of neutrons, protons and low-mass fragments in \secref{modelvalid}.
In~\secref{ddse} the extracted constraints for $L$ and $K_{sym}$ parameters are presented. A detailed
investigation of possible residual model dependences and systematical uncertainties of the obtained
results is performed. Differences with respect to constraints extracted using a previous version
of the model~\cite{Cozma:2013sja} are explained in detail. The obtained results are compared with
existing constraints for the isospin dependent component of nuclear matter compressibility and recommendations for future
experimental measurements of flow ratios are presented in \secref{discussion}. The article ends with a section 
devoted to summary and conclusions.

\section{The model}
\seclab{model}
\subsection{Transport model}
\seclab{transport}

Heavy-ion collision dynamics is simulated using an upgraded version~\cite{Cozma:2014yna,Cozma:2016qej}
of the T\"ubingen quantum molecular dynamics model (QMD) transport model~\cite{Khoa:1992zz,UmaMaheswari:1997ig}
which provides a semi-classical framework for the description of such reactions and accounts for relevant
quantum aspects such as  stochastic scattering and Pauli blocking of nucleons. It includes the production of
all nucleonic resonances with masses below 2 GeV, in total 11 N$^*$ and 10 $\Delta$ resonances.

QMD-type transport models provide a solution for the time dependence of the density matrix of the system by the
method of the Weyl transformation applied to the many-body Schr\"odinger equation. 
Generally, the expectation values for the position and momentum operators can be shown to satisfy the classical
Hamiltonian equations of motion ~\cite{deGroot:1972aa,Hartnack:1997ez}.
These can be factorized to each particle by approximating the total wave-function of the system
as the product of individual nucleon wave functions, represented by Gaussian wave packets of
finite spread in phase space,
\begin{eqnarray}
\frac{d\vec{r}_i}{dt}=\frac{\partial \langle U_i \rangle}{\partial \vec{p}_i}+\frac{\vec{p}_i}{m},\qquad
\frac{d\vec{p}_i}{dt}=-\frac{\partial \langle U_i \rangle}{\partial \vec{r}_i}\,.
\end{eqnarray}
The average of the potential operator is understood to be taken over the entire phase-space
and weighted by the Wigner distribution of particle $i$. The potential operator $U_i$ is in
this case the sum of the Coulomb and strong interaction potential operators.
In all kinematic equations the relativistic relation between mass, energy and momentum is used.

Description of pion production in heavy-ion collisions close to threshold requires transport models
that enforce the conservation of the energy of the system (locally or globally) by taking into
account the potential energies of hadrons in nuclear matter
~\cite{Cozma:2014yna,Cozma:2016qej,Song:2015hua,Zhang:2017mps}. In the present model this is achieved
by including potential energies in the total energy conservation constraint
imposed when determining the final state of a 2-body scattering, decay or absorption process,
\begin{eqnarray}
 \sum_j \sqrt{p_j^2+m_j^2}+U_j=\sum_i \sqrt{p_i^2+m_i^2}+U_i,
\end{eqnarray}
both indexes running over all particles present in the system and corresponding, from left to right, 
to the final and initial states of an elementary reaction. This scenario has been referred to as the
``global energy conservation'' (GEC) scenario in~\cite{Cozma:2014yna}.
Their impact on flow ratio observables has been shown to be within the uncertainty induced by
the experimental data~\cite{Cozma:2014yna}. Most of the results presented in this article have therefore
been obtained by neglecting these effects, corresponding to the standard so-called ``vacuum energy
conservation''(VEC) scenario in the mentioned reference. It is nevertheless important to present results also
for this scenario in order to asses the impact on the curvature parameter $K_{sym}$ and to confirm
compatibility of the extracted constraints with a similar study of pionic observables, in which case the conservation
of the total energy of the system is crucial for a faithful description of experimental data~\cite{Cozma:2016qej}.

\subsection{Initialization of nuclei}
\seclab{ininuc}
Following the first findings the Code Comparison Project \cite{Xu:2016lue}, the initialization part
of the TuQMD model has been modified to better reproduce nuclear density profiles~\cite{Cozma:2016qej}.
In previous versions of the model~\cite{Cozma:2011nr,Cozma:2013sja,Cozma:2014yna},
the radius mean square (rms) of initialized nuclei was determined 
solely from the position of the centroids of the wave function of nucleons. This is however inaccurate
for the case of Gaussian-type nucleon  wave functions of finite width, as used in QMD transport models,
leading to an effective larger rms. The appropriate expression reads
\begin{eqnarray}
 \langle r^2 \rangle=\frac{1}{N}\sum_{i=1}^{N} (\langle \vec r \rangle - \vec r_i)^2+\frac{3}{2}L_N,
\eqlab{correctedrms}
\end{eqnarray}
where $L_N$ is the square of the nucleon wave function width, the used convention for
the parametrization of the nucleon wave function being the same as in
Ref.~\cite{Hartnack:1997ez}. The difference between the previously used and the
appropriate value grows with increasing wave function width, reaching about 10$\%$
for values customarily used in transport models in connection
with heavy nuclei. Additionally the distribution used to sample nucleon centroids was not
leading to good enough density profiles and had to be fine tuned to the nucleus of interest.

\begin{figure}
\begin{center}
\resizebox{0.40\textwidth}{!}{
\includegraphics{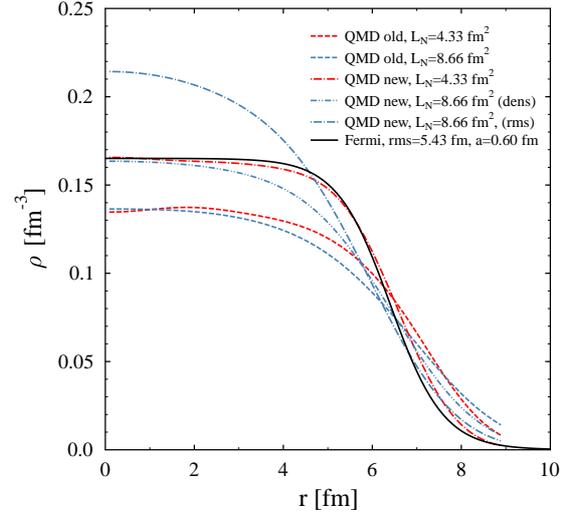}}
\end{center}
\caption{Nucleon density profiles for the $^{197}$Au nucleus produced by the initialization
routines of the transport model used in Ref.~\cite{Cozma:2011nr,Cozma:2013sja,Cozma:2014yna} (``QMD old'')
and that of Ref.~\cite{Cozma:2016qej} and present version (``QMD new'') for two values
of the wave function width $L_N$ that correspond to those customarily employed in the literature for
light ($L_N$=4.33 fm$^2$) and heavy ($L_N$=8.66 fm$^2$) nuclei~\cite{Hartnack:1997ez}. For the higher shown value of $L_N$ it is impossible
to describe the value of the central density and rms simultaneously. Consequently two cases are shown that
were built to describe only one of these quantities. The realistic density profile~\cite{Angeli:2013aa,DeJager:1987qc}
corresponding to a rms value of $\sqrt{<r^2>}$=5.43 fm and diffuseness a=0.6 fm is also shown.}
\figlab{densprofile}
\end{figure}

In~\figref{densprofile} density profiles for the $^{197}$Au nucleus as produced by different versions
of the model (``QMD old'' and ``QMD new'') and for different values of the wave function width $L_N$ are shown.
Most of the difference between the new and old versions is due to the $L_N$ dependent correction term in
the expression for the rms,~\eqref{correctedrms}. For the larger value $L_N$=8.66 fm$^2$, 
quoted in the literature in the context of achieving better stability for heavy nuclei~\cite{Hartnack:1997ez}, 
a good enough description of the realistic density profile cannot be attained.
This can be traced back to the distribution function of nucleon centroids becoming significantly negative
in the skin region. 

The choice $L_N$=4.33 fm$^2$ exhibits the same problem, its severity is however much smaller. Consequently
a reasonable description of the realistic density profiles can be achieved in this case, see~\figref{densprofile}.
The choice of the value for the wave function width $L_N$ amounts to a compromise between
a good description of empirical density profiles and high stability of the initialized nucleus, as they
cannot be achieved simultaneously. The former is without argument more important for the study of elliptic flow
observables for impact energies above 100 MeV/nucleon. Consequently, we use $L_N$=4.33 fm$^{2}$ through out this study.

Neutrons and protons are initialized using the same coordinate-space probability distribution.
Consequently, possible neutron skin effects are neglected in this study. Momenta of nucleons are initialized randomly
by taking into account the value of the local Fermi momentum with the constraint that the nucleon in question is
bound and that it is not Pauli blocked. Only initializations of nuclei with a binding energy per nucleon in the range
7.0 MeV $\leq E_B \leq $ 9.0 MeV are accepted.

The impact of the improvement of density profiles on pion multiplicities in central collisions is small, 
leaving the results of Ref.~\cite{Cozma:2014yna} unchanged. In contrast the impact on flow observables 
in mid-central and, especially, peripheral collisions is non-negligible as it will be shown in~\secref{prevmodel}.

\subsection{Pauli blocking}
The Pauli blocking algorithm is one of the sources of model dependence of transport models~\cite{Xu:2016lue}
as there is no universally accepted method for the calculation of occupancy probability once the unavoidable
approximations that make transport models solvable are introduced and the fermionic nature of nucleons is lost.
The wide variety of Pauli blocking algorithms implemented in existing transport models lead to Pauli blocking factors
that can differ even by a factor of two~\cite{Xu:2016lue}.

The Pauli blocking algorithm implemented in TuQMD is the same as that of an earlier version of
the QMD model~\cite{Aichelin:1991xy}. Every potential two-nucleon collision is blocked with a probability
\begin{eqnarray}
 P_{blocked}=1-(1-P_1)(1-P_2),
\end{eqnarray}
where $P_1$ and $P_2$ are the occupation fractions of phase-space around nucleons 1 and 2 respectively.

The determination of the occupation fraction makes use of the property of the Wigner distribution (the Weyl
transform of the wave function) of having a magnitude less than $(2/h)^3$ in absolute terms
~\cite{deGroot:1972aa}. This implies that it must take values different from zero in a volume 
in phase-larger larger than $(h/2)^3$. Consequently the value $V_0$=2$(h/2)^3$ is adopted, the factor of 2 accounting
for spin degeneracy. Two spheres of radii $R_r$ and $R_p$ are associated to each nucleon in
$r$- and $p$-space respectively. The occupancy fraction is determined as the ratio between the sum
of volumes of spheres overlaps of the nucleon of interest with nearby nucleons and the volume $V_0$
\begin{eqnarray}
 P_i^{(1)}=2\sum_j \delta_{\tau_i \tau_j}\frac{V_{ij}^{(r)} V_{ij}^{(p)}}{V_0}\,.
\end{eqnarray}
The isospin dependence of the occupancy fraction is introduced by adding contributions only from
nucleons $j$ with the same isospin as that of nucleon $i$ ($\tau_i=\tau_j$). The factor of 2 in front of the
sum ensures that in the case of isospin symmetric matter similar blocking probabilities as those of the original
isospin independent TuQMD Pauli blocking algorithm are recovered. More importantly, with this choice 
the initial Fermi distribution of nucleon momenta are preserved for durations larger than 200 fm/c when
isospin symmetric nuclear matter is simulated in a box. Setting $R_r$=3.0~fm, the value of $R_p$ is determined from the relation
\begin{eqnarray}
 V_0=\frac{4\pi R_r^3}{3}\,\frac{4\pi R_p^3}{3}\,.
\end{eqnarray}
For the case of a nucleon close to the surface only part of the classically available phase-space is allowed.
To correct for this effect the occupancy fraction due to nucleons that find themselves in a direction, both
in coordinate and momentum space, within a cone of 45$^\circ$ opening angle around the direction nucleon-center of mass and
nucleon momentum - center of mass momentum respectively. The location of the center of mass and the total momentum
are determined considering only nucleons for which there is a non-zero overlap of the phase-space spheres with
those of the nucleon in question. The result is extrapolated to 4$\pi$ solid angle by multiplying it with
the factor $\frac{4}{2-\sqrt{2}}\approx 6.83$ and denoted $P_i^{(2)}$. The final occupancy fraction is determined as
\begin{eqnarray}
 P_i= Max \{P_i^{(1)},P_i^{(2)}\}\,.
\end{eqnarray}
This algorithm has been developed at times when the evaluation of transcendental functions was
computationally expensive and misses contributions from nucleons further away in $r$ and $p$-space
than 2$R_r$ and 2$R_p$ respectively. This approximation grows worse with increasing wave function width $L_N$.

To correct for this, a second Pauli algorithm, that takes into account the proper
overlap of the gaussian Wigner functions associated to each nucleon, has been implemented. 
Starting from the expression for the Wigner function of a nucleon of momentum $\vec {p}_i$ 
and centroid position $\vec {r}_i$,
\begin{eqnarray}
 f_i(\vec{r},\vec{p})=\frac{1}{\pi^3 \hbar^3}\,e^{-(\vec{r}-\vec{r}_i)^2\,\frac{2}{L_N}}
 \,e^{-(\vec{p}-\vec{p}_i)^2\,\frac{L_N}{2\hbar^2}}\,,
\end{eqnarray}
the occupancy fraction of the phase-space of nucleon $i$ by that of nucleon $j$ is readily found to be equal to
\begin{eqnarray}
 P_{ij}=\,Erfc\bigg(\frac{|\vec{r}_i-\vec{r}_j|}{\sqrt{2\,L_N}}\bigg)\,
 Erfc\bigg(\frac{|\vec{p}_i-\vec{p}_j|}{4}\,\frac{\sqrt{2\,L_N}}{\hbar}\bigg),
\end{eqnarray}
with $Erfc(x)$ being the complementary error function. In this case the total occupancy fraction of nucleon $i$
is defined to be
\begin{eqnarray}
 P_i=\frac{1}{2}\,\sum_j\,\delta_{\tau_i\tau_j}P_{ij}\,,
\end{eqnarray}
omitting thus possible surface corrections. The factor $1/2$ takes into account spin degeneracy.
For the case of central Au+Au collisions at an impact energy of 400 MeV/nucleon
the new blocking algorithm leads to the decrease of the number of successful collisions by about 15$\%$ as
compared to the original one. The two Pauli blocking algorithms (PBA) will be referred in the following as
 sPBA (``spheres'') and gPBA (``gaussians'') respectively.

%In the vicinity of nucleus's surface a correction is performed such the only the classically allowed portion
%of the phase-space is taken into account. Baryonic resonances are considered to be unaffected by Pauli blocking.
\subsection{Equation of state}
\label{sec:eos}

A modified version of the MDI Gogny-inspired parametrization of the equation of state of
nuclear matter~\cite{Das:2002fr}, named MDI2 in the following, has been selected for the present study. Its potential component reads
\begin{eqnarray}
\eqlab{eos}
\frac{E}{N}(\rho,\beta)&=&A_u(x,y)\frac{\rho(1-\beta^2)}{4\rho_0}+A_l(x,y)\frac{\rho(1+\beta^2)}{4\rho_0} \\
&&+\frac{B}{\sigma+1}\frac{\rho^{\sigma}}{\rho_0^\sigma}\,(1-x\beta^2)+
\frac{D}{3}\frac{\rho^2}{\rho_0^2}\,(1-y\beta^2) \nonumber\\
&&+\frac{1}{\rho\rho_0}\sum_{\tau,\tau'} C_{\tau \tau'}\!\!\int\!\!\int d^{\!\:3} \vec{p}\,d^{\!\:3} \vec{p}\!\;'
\frac{f_\tau(\vec{r},\vec{p}) f_{\tau'}(\vec{r},\vec{p}\!\;')}{1+(\vec{p}-\vec{p}\!\;')^2/\Lambda^2}. \nonumber
\end{eqnarray}
The difference resides in the extra term proportional to the $D$ parameter, that has been introduced
in order to allow independent variations of the slope $L$ and curvature $K_{sym}$ parameters of the SE, while
keeping the neutron-proton isovector effective mass difference constant. This is a mandatory feature given
the present uncertainties on  the value of this parameter and its potential impact on the extracted constraints
on the SE stiffness. A similar modified Gogny-inspired parametrization of the EoS has
been recently used to study thermal properties of asymmetric nuclear matter~\cite{Xu:2014cwa}, however in this case 
the extra parameters introduced lead to correlated modifications of the slope $L$, curvature  $K_{sym}$ and isovector
effective mass difference. 

%The strategy employed in Ref.~\cite{Xu:2014cwa} to fix the parameters of the momentum dependent 
%part of the interaction has been adopted in this study as well.

The corresponding single-particle potential is given by

\begin{eqnarray}
\eqlab{sympot}
 U_\tau(\rho,\beta,p)&=&A_u(x,y)\frac{\rho_{\tau'}}{\rho_0}+A_l(x,y)\frac{\rho_{\tau}}{\rho_0} \\
&&+B\,\Big(\frac{\rho}{\rho_0}\Big)^\sigma(1-x\beta^2)
-4\tau x\frac{B}{\sigma+1}\frac{\rho^{\sigma-1}}{\rho_0^\sigma}\beta\rho_{\tau'} \nonumber\\
&&+D\,\Big(\frac{\rho}{\rho_0}\Big)^2(1-y\beta^2)
-4\tau y\frac{D}{3}\frac{\rho}{\rho_0^2}\beta\rho_{\tau'} \nonumber\\
&&+\frac{2C_{\tau \tau}}{\rho_0}\int d^{\!\:3} \vec{p}\!\;'\, \frac{f_\tau(\vec{r},\vec{p}\!\;')}{1+(\vec{p}-\vec{p}\!\;')^2/\Lambda^2} \nonumber\\
&&+\frac{2C_{\tau \tau'}}{\rho_0}\int d^{\!\:3} \vec{p}\!\;'\, \frac{f_{\tau'}(\vec{r},\vec{p}\!\;')}{1+(\vec{p}-\vec{p}\!\;')^2/\Lambda^2}. \nonumber
\end{eqnarray}
In the above expressions $\rho$, $\beta$ and $p$ denote the density, isospin asymmetry and momentum variables respectively.
The label $\tau$ designates the isospin component of the nucleon or resonance and takes the value $\tau$=+1 (-1) 
for neutrons (protons). For cold nuclear matter it holds $f_\tau(\vec{r},\vec{p})=(2/h^3)\Theta(p_F^\tau-p)$, with $p_F^\tau$ the 
Fermi momentum of nucleons with isospin $\tau$.

The dependence of the $A_u$ and $A_l$ parameters on the stiffness parameters $x$ and $y$ is
required to  be such as to lead to an expression of the symmetry energy that is independent of them for
a particular value of the density, denoted here as $\tilde\rho$. This can be chosen to be different from
the saturation density $\rho_0$ as opposed to Ref.~\cite{Das:2002fr} with
\begin{eqnarray}
\eqlab{modpardef}
 A_u(x,y)&=&A_u^0-\frac{2B(x-1)}{\sigma+1}\frac{\tilde\rho^{\sigma-1}}{\rho_0^{\sigma-1}}-\frac{2D(y-1)}{3}\frac{\tilde\rho}{\rho_0} \nonumber\\
&=&\tilde {A_u}-\frac{2xB}{\sigma+1} \frac{\tilde\rho^{\sigma-1}}{\rho_0^{\sigma-1}}-\frac{2yD}{3}\frac{\tilde\rho}{\rho_0}  \\
A_l(x,y)&=&A_l^0+\frac{2B(x-1)}{\sigma+1}\frac{\tilde\rho^{\sigma-1}}{\rho_0^{\sigma-1}}+\frac{2D(y-1)}{3}\frac{\tilde\rho}{\rho_0} \nonumber\\
&=&\tilde {A_l}+\frac{2xB}{\sigma+1} \frac{\tilde\rho^{\sigma-1}}{\rho_0^{\sigma-1}}+\frac{2yD}{3}\frac{\tilde\rho}{\rho_0},
\end{eqnarray}
with obvious definitions for $\tilde {A_u}$ and $\tilde {A_l}$ that include all terms independent of $x$ or $y$.

The analytical expressions for the integrals appearing in \eqref{eos} and \eqref{sympot} and the coefficients of
their Taylor expansion in isospin asymmetry are needed in the process of fixing model parameters. 
%While the derivation can be performed using only elementary methods, it is rather tedious.
They are listed, for convenience, in Appendix A.

Consistency with the transport model requires the use of the relativistic expression for kinetic energy
in the process of fixing the parameters of the potential. The contribution of the kinetic term to the
equation of state is given by
\begin{eqnarray}
 \frac{E^{kin}}{N}(\rho,\beta)&=&\sum_{\tau=-1,1} \frac{3\,(1+\tau \beta)}{2\,p_F^3(\tau)}\,
 \Bigg[\,\frac{p_F^3(\tau)}{4}\sqrt{p_F^2(\tau)+m^2} \nonumber\\
 &&+\frac{m\,p_F^2(\tau)}{8}\sqrt{p_F^2(\tau)+m^2} \\
 &&-\frac{m^4}{8} \ln{\frac{m+\sqrt{p_F^2(\tau)+m^2}}{m}}\,\Bigg].\nonumber
\end{eqnarray}
Expanding it in powers of the isospin asymmetry $\beta$, the contribution of the kinetic
term to the symmetry energy is found to take a simple form
\begin{eqnarray}
 S^{kin}(\rho,\beta)&=&\frac{p_F^2}{6\sqrt{p_F^2+m^2}}.
\end{eqnarray}
As before, $p_F$ is the Fermi momentum of symmetric nuclear matter of density $\rho$.
Also the non-relativistic expressions for the effective isoscalar mass and the isovector neutron-proton
mass difference have to be replaced by their relativistic counterparts. For the former the following
expression is found
\begin{eqnarray}
m^*_s\,^2(\rho,p)=\frac{m^2-2Ep\frac{\partial U}{\partial p}+E^2\big(\frac{\partial U}{\partial p}\big)^2}
{\big[ 1+\frac{E}{p}\,\frac{\partial U}{\partial p} \big]^2}.
\end{eqnarray}
The expression of the latter can then be derived in a straightforward manner. For the
case when second and higher powers of $\partial U/\partial p$ can be neglected one arrives at an approximation that
resembles the corresponding non-relativistic one~\cite{Xu:2010fh,Li:2013ola}, with the exception of kinematical factors
\begin{eqnarray}
 \delta m^*_{n-p}(\rho,\beta,p)&\equiv&\frac{m^*_n-m^*_p}{m}\\
 &\approx& \frac{\frac{E}{p} \big(1+\frac{p^2}{2m^2}\big) 
 \big( \frac{\partial U_p}{\partial p}- \frac{\partial U_n}{\partial p}\big)}{1+\frac{E}{p}
 \big( \frac{\partial U_p}{\partial p}+ \frac{\partial U_n}{\partial p}\big)}.\nonumber
\end{eqnarray}
In the above relations, the relativistic energy is denoted by $E=\sqrt{p^2+m^2}$, while $U$, $U_n$ and $U_p$
stand for the isoscalar, neutron and proton single-particle potentials of \eqref{sympot}, respectively . In the actual calculations
the exact relativistic expressions for effective masses have been used.

With all the needed ingredients in place the parameters appearing in the potential part of the EoS, \eqref{eos},
can be fixed. The 11 unknown parameters $\Lambda$, $C_l=C_{1,1}=C_{-1,-1}$, $C_u=C_{1,-1}=C_{-1,1}$, $\tilde{A}_l$, 
$\tilde{A}_u$, $B$, $\sigma$, $D$, $\tilde{\rho}$, $x$ and $y$ are determined from a non-linear system of equations that use as
input the following quantities: value of the optical potential at infinite momentum, 
effective nucleon mass in isospin symmetric matter, isovector effective mass difference, value of the saturation
density of isospin symmetric matter, value of the binding energy at saturation, compressibility $K_0$
and skewness $J_0$ parameters of isospin symmetric matter, value of the symmetry energy at density $\tilde{\rho}$ 
and the slope $L$ and curvature $K_{sym}$ of symmetry energy at saturation. The choice made for each of these 
quantities will be discussed in the following.

The values of the $\Lambda$, $C_l$ and  $C_u$ parameters are determined by optimally reproducing the momentum
dependent part of the optical potential and the value of neutron-proton effective mass difference. 
The first constraint is compatible with an effective isoscalar nucleon mass $m_s^*$=0.7$m$ for both the
Gogny inspired MDI interaction~\cite{Das:2002fr} and the empirical nucleon optical potentials
~\cite{Arnold:1982rf,Hama:1990vr,Cooper:1993nx}. The energy dependence of these two classes
of potentials is however different above 200 MeV/nucleon kinetic energy, the former being attractive while latter is
repulsive. Given the well known impact of the momentum dependent part of the optical potential on heavy-ion dynamics
in general and flow observables in particular, model parameters were fixed such as to reproduce as closely
as possible the momentum dependence of the empirical optical potential. To that end the value of the optical
potential at infinite incident momentum is required to be $U_\tau(\rho_0,0,\infty)$=75 MeV~\cite{Xu:2014cwa}. This
choice leads to a good description of the empirical energy dependence of the optical potential, small deviations from it
being visible only at very low momenta. It has been verified that these 
imperfections impact the value of elliptic flow ratios only marginally. Alternative choices, such as requiring 
the value of the empirical potential at low kinetic energies to be reproduced, lead to potentials that resemble 
the energy dependence of the MDI interaction and thus deviate strongly from the empirical ones above 200 MeV/nucleon
kinetic energy.

The extraction of the value of the neutron-proton effective mass difference from experimental data has been
the aim of several recent investigations. For the present study a value at saturation density, 
$\delta m^*_{n-p}$($\rho_0,\beta,p=p_F$)=0.33$\beta$, in close agreement with the average of currently undisputed 
results~\cite{Xu:2010fh,Li:2013ola,Li:2014qta,Zhang:2015qdp}, has been adopted. The precise value of this
quantity is however still the subject of ongoing debates spurred by a few controversial lower (or even of opposite
sign) extracted values~\cite{Coupland:2014gya,Kong:2017nil}, that still require
further confirmation or possibility alternative explanations. Consequently the impact of this quantity on the
extracted constraints for the stiffness for the SE will be studied by varying its magnitude between the
lowest and highest allowed values, at 1$\sigma$ level, as reported in these studies.

\begin{table}
\caption{Input quantities and their values (first and second columns) together with the model
parameters appearing in~\eqref{eos} and their determined values (third and fourth columns). }
\tablab{model_input_params}       % Give a unique label
\begin{center}
\begin{tabular}{cr@{.}l|cr@{.}l}
\hline\noalign{\smallskip}
\multicolumn{3}{c|}{Input} & \multicolumn{3}{c}{Parameters}\\
\noalign{\smallskip}\hline\noalign{\smallskip}
$\rho_0$ [fm$^{-3}$] & 0&16 & $\Lambda$ [MeV] & 708&001\\
$E_B$ [MeV] & -16&0 & $C_l$ [MeV] & -13&183\\
$m_s^*/m$ & 0&70 & $C_u$ [MeV] & -140&405 \\
$\delta_{n-p}^*$ ($\rho_0, \beta=$0.5) & 0&165 & $B$ [MeV] & 137&305 \\
$K_0$ [MeV] & 245&0 & $\sigma$  &  1&2516 \\
$J_0$ [MeV] & -350&0 & $\tilde{A_l}$ [MeV] & -130&495 \\
$\tilde{\rho}$ [fm$^{-3}$] & 0&10 & $\tilde{A_u}$ [MeV] & -8&828 \\
S($\tilde{\rho}$) [MeV] & 25&5 & $D$ [MeV] & 7&357 \\
\noalign{\smallskip}\hline
\end{tabular}
\end{center}
%\vspace*{2cm}  % with the correct table height
\end{table}

\begin{table}
\caption{Values of the $x$ and $y$ parameters for selected values for $L$ and $K_{sym}$. The other
parameters of the model take the values listed in~\tabref{model_input_params}. In the upper half
$L$ and $K_{sym}$ combinations corresponding to the MDI~\cite{Das:2002fr} (for the listed values of $x_{MDI}$)
and cMDI2 potentials are presented. }
\tablab{lsymxyvals}
\begin{center}
\begin{tabular}{r@{.}lr@{.}l|r@{.}lr@{.}l|c}
\hline\noalign{\smallskip}
\multicolumn{4}{c|}{Input} & \multicolumn{5}{c}{Parameters}\\
\noalign{\smallskip}\hline\noalign{\smallskip} 
\multicolumn{2}{c}{$L$ [MeV]} & \multicolumn{2}{c|}{$K_{sym}$ [MeV]} 
& \multicolumn{2}{c}{$x$} & \multicolumn{2}{c|}{$y$} & \multicolumn{1}{c}{$x_{MDI}$}\\
\noalign{\smallskip}\hline\noalign{\smallskip}
151&0 &349&0  &1&179  &-14&459 & -2.0\\
106&0 & 135&0 & 1&028 & -9&021 & -1.0\\
60&5 & -81&0 & 0&879 & -3&543 & \phantom{-}0.0\\
15&0 & -298&0 & 0&721 & 1&990 & \phantom{-}1.0\\
-31&0 & -512&0 & 0&571 & 7&429 & \phantom{-}2.0\\
\noalign{\smallskip}\hline\noalign{\smallskip}
60&0 & 600&0 & 6&715 & -41&826 &\\
60&0 & 300&0 & 4&152 & -24&994 &\\
60&0 &   0&0 & 1&589 & -8&161 & N/A\\
60&0 & -300&0 & -0&973 & 8&672 &\\
60&0 & -600&0 & -3&536 & 25&504 &\\
%0&0 &  0&0 &  3&826 & -16&920 \\
%20&0 & 0&0 & 3&080  &  -14&000\\
%40&0 & 0&0 & 2&335  & -11&081 \\
%60&0 & 0&0 & 1&589  &  -8&161\\
%80&0 & 0&0 &  0&844 &  -5&241\\
%100&0 & 0&0 & 0&098  & -2&322\\
%120&0 & 0&0 & -0&647  &  0&598\\
%140&0 & 0&0 &  -1&393 &  3&517\\
\noalign{\smallskip}\hline
\end{tabular}
\end{center}
%\vspace*{3cm}  % with the correct table height
\end{table}

Four parameters, $\tilde {A_u}+\tilde {A_l}$, $\sigma$, $B$ and $D$ are determined from the chosen
density dependence of the EoS of symmetric nuclear matter, namely saturation density $\rho_0$=0.16 fm$^{-3}$,
binding energy per nucleon at saturation $E_B$=-16 MeV, the values of the incompressibility modulus $K_0$ and
skewness parameter $J_0$. The value of the third one is set to $K_0$=245 MeV,
in agreement with recent extractions from nuclear structure and heavy-ion collision experimental
data~\cite{Danielewicz:2002pu,Colo:2004mj,Wang:2013wca,Fevre:2015fza}. The value of the skewness parameter
is not known accurately at present. Its extraction from 
earth-based laboratory measurements~\cite{Farine:1997vuz,Cai:2014kya,Meixner:2013ava} 
has been impacted up to now by the same potentially unphysical correlations with the lower 
order Taylor coefficients~\cite{Chen:2011ib} as for the case of the symmetry energy parameters.
Determination from astrophysical observables is still affected by large uncertainties~\cite{Steiner:2010fz}. 
Due to the lack of more trustworthy information, the value $J_0$=-350 MeV has been chosen for this study,
in agreement with the ranges put forward in the mentioned references. This choice
for $K_0$ and $J_0$ reproduces closely the density dependence of the nuclear matter incompressibility $M_c$
around the so-called crossing density $\rho_c$=0.10 fm$^{-3}$ \cite{Khan:2012ps,Khan:2013mga}.
The stiffness of the isoscalar equation of state has an important impact on the magnitude of collective flows.
By taking the difference or ratio of observables corresponding to isospin partners this sensitivity is greatly
suppressed~\cite{Cozma:2011nr}. Consequently, uncertainties in the isoscalar EoS will have a limited, though finite, impact
on the extraction of the slope and curvature of the symmetry energy.

The remaining four parameters, $\tilde {A_u}-\tilde {A_l}$, $\tilde{\rho}$, $x$ and $y$ determine the density dependence
of the symmetry energy. The parameter $\tilde{\rho}$ defines the density at which the magnitude of the symmetry energy
is supposed to be known from other sources and used as input to the model. In the context of a Taylor
expansion of the symmetry energy around the saturation point it is natural and customary to make 
the choice $\tilde{\rho}$=$\rho_0$. The value of the symmetry energy at saturation is 
however not accurately known at present. In contrast, it has been possible to extract precise values at
sub-saturation densities from experimental data of static properties of nuclei. The study of properties
of doubly-magic nuclei (binding energies, rms charge radii and single-particle energies) using a few carefully selected
Skyrme energy density functionals has resulted in a value for the symmetry 
energy $S(\tilde\rho)$=25.5$\pm$1.0 MeV at $\tilde\rho$=0.10 fm$^{-3}$~\cite{Brown:2013mga}. 
Similarly, the binding energy difference of heavy isotope pairs has allowed the extraction of an even more
precise value $S(\tilde\rho)$= 26.65$\pm$0.22 MeV at a slightly higher value for the density
$\tilde\rho$=0.11 fm$^{-3}$~\cite{Zhang:2013wna}. These empirical findings are in good agreement
with many body calculations of the neutron matter EoS that use as input microscopical N$^3$LO chiral
perturbation theory effective potentials~\cite{Kruger:2013kua,Drischler:2015eba,Drischler:2016djf}.
The empirical value extracted in Ref.~\cite{Brown:2013mga} will be used as input to the model. As it will
be shown in the results section, elliptic flow ratios at impact energies used in this study are rather insensitive
to uncertainties associated to this quantity.

The values of input quantities and majority of model parameters appearing in~\eqref{eos} are summarized
in \tabref{model_input_params}. Values for $x$ and $y$ model parameters corresponding to selected
combinations for $L$ and $K_{sym}$ are listed in \tabref{lsymxyvals}. The combinations in the upper half of the table
correspond to $L$ and $K_{sym}$ values that the MDI potential~\cite{Das:2002fr} leads to for integer values of $x$
(denoted $x_{MDI}$ in \tabref{lsymxyvals}) and that have been used in previous studies
\cite{Cozma:2011nr,Cozma:2013sja,Cozma:2014yna,Cozma:2016qej}. They will be mimicked in the present study by
a constrained version of the full MDI2 potential referred to as cMDI2 in \secref{ddse}.

\subsection{Coalescence model}
\seclab{coalescence}
A minimum spanning-tree (MST) algorithm is employed to generate the final state spectra of intermediate
energy heavy-ion collisions. Within such a model, nucleons that are located closer than
a predefined range in both coordinate and momentum space are assumed to be part of a cluster.
To take into account possible isospin effects and alleviate some of the shortcoming of semi-classical
transport models that lead to underprediction of light fragment multiplicities,
three cut-off parameters, that fix the maximum
allowed separation of a nucleon belonging to a fragment from nucleons of the same cluster,
have been introduced in coordinate space, $\delta r_{nn}$, $\delta r_{np}$ and $\delta r_{pp}$. 
In momentum space a nucleon is considered to belong to a cluster if its momentum in the rest frame
of the (potential) cluster is smaller than a certain cut-off value. Consequently,
two cut-off parameters, $\delta p_n$ and $\delta p_p$ have been considered. To avoid the possibility
of final spectra depending on the sequence in which nucleons are tested in the coalescence algorithm
the momentum cut-off parameters are increased in steps from 0 to the desired value~\cite{Neubert:2003qm}.
Within this approach clusters identified at an earlier step are used as input for the following one
and can grow or evaporate nucleons depending on the momenta of the extra nucleons that may fulfill
the coalescence criterion at the current step. This procedure also leads to clusters with nucleons having minimum momenta
in the cluster's rest frame and thus reduce the dependence of the final spectra on the moment at which
the coalescence algorithm is applied. In the actual calculations a number of steps equal to 5 has been used.

The coalescence algorithm is applied to identify all clusters with A$\leq$15. Also a
number of 23 additional clusters with A$>$15 that correspond to known stable or unstable
isotopes of $B, C, N$ and $O$ are identified. Clusters with lifetimes larger than 
1 ms~\cite{LundNDS:2017aa,Nudat2:2017aa} are considered
as stable while the rest are decayed until clusters stable against strong interaction decays are reached
(ex. $^4$Li$\rightarrow$p+$^3$He). For clusters with A$\leq$15 that do not correspond to a known
stable or unstable isotope, protons or neutrons are evaporated for proton and neutron rich clusters
respectively until a known nucleus is reached (ex. 1p8n$\rightarrow$6n+$^3$H). Nucleons 
belonging to clusters with other masses are discarded from final spectra.

\begin{table}
\caption{Values of $r$-space coalescence parameters determined from a fit of FOPI experimental
multiplicities of free nucleons and light clusters in central Au+Au collisions (b$\le$ 2 fm) at 400 MeV/nucleon~\cite{Reisdorf:2010aa}
for three values of $\delta$p. The asy-EoS parameters have been set to $L$=80.0 MeV and $K_{sym}$=0.0 MeV.}
\tablab{rcoalpars}
\begin{center}
\begin{tabular}{c|c|c|c}
\hline\noalign{\smallskip}
$\delta$p [GeV/c] & $\delta$r$_{pp}$ [fm] & $\delta$r$_{np}$ [fm] &$\delta$r$_{nn}$ [fm]\\
\noalign{\smallskip}\hline\noalign{\smallskip}
0.15 & 6.25 & 6.78 & 7.03 \\
0.20 & 3.75 & 4.22 & 4.25 \\
0.25 & 2.25 & 3.25 & 2.75 \\
\noalign{\smallskip}\hline
\end{tabular}
\end{center}
\end{table}

\begin{figure*}[ht]
%\begin{minipage}{0.45\textwidth}
\begin{center}
\resizebox{0.45\textwidth}{!}{
\includegraphics{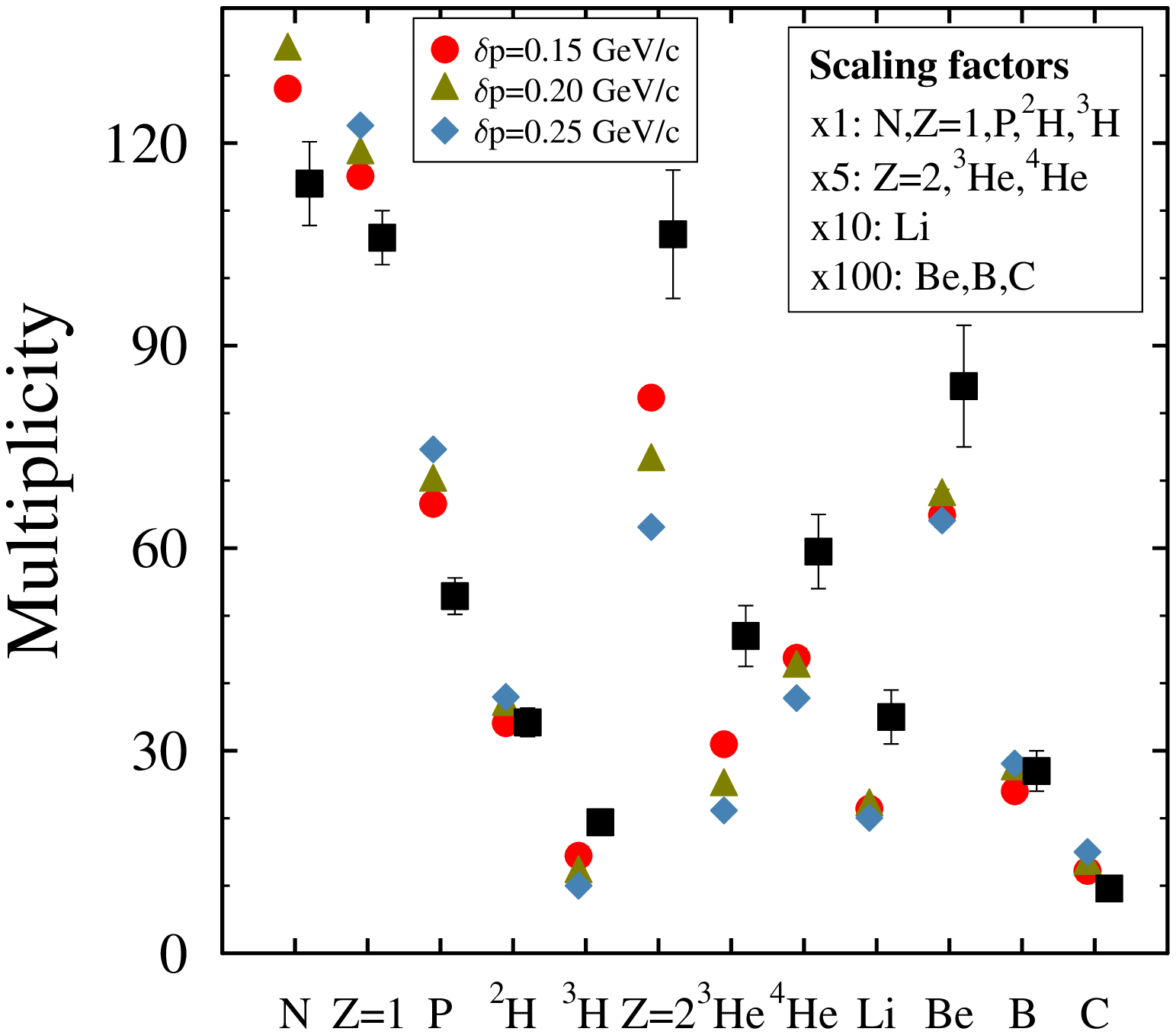}}
%\end{minipage}{0.45\textwidth}
%\begin{minipage}{0.45\textwidth}
\resizebox{0.45\textwidth}{!}{
\includegraphics{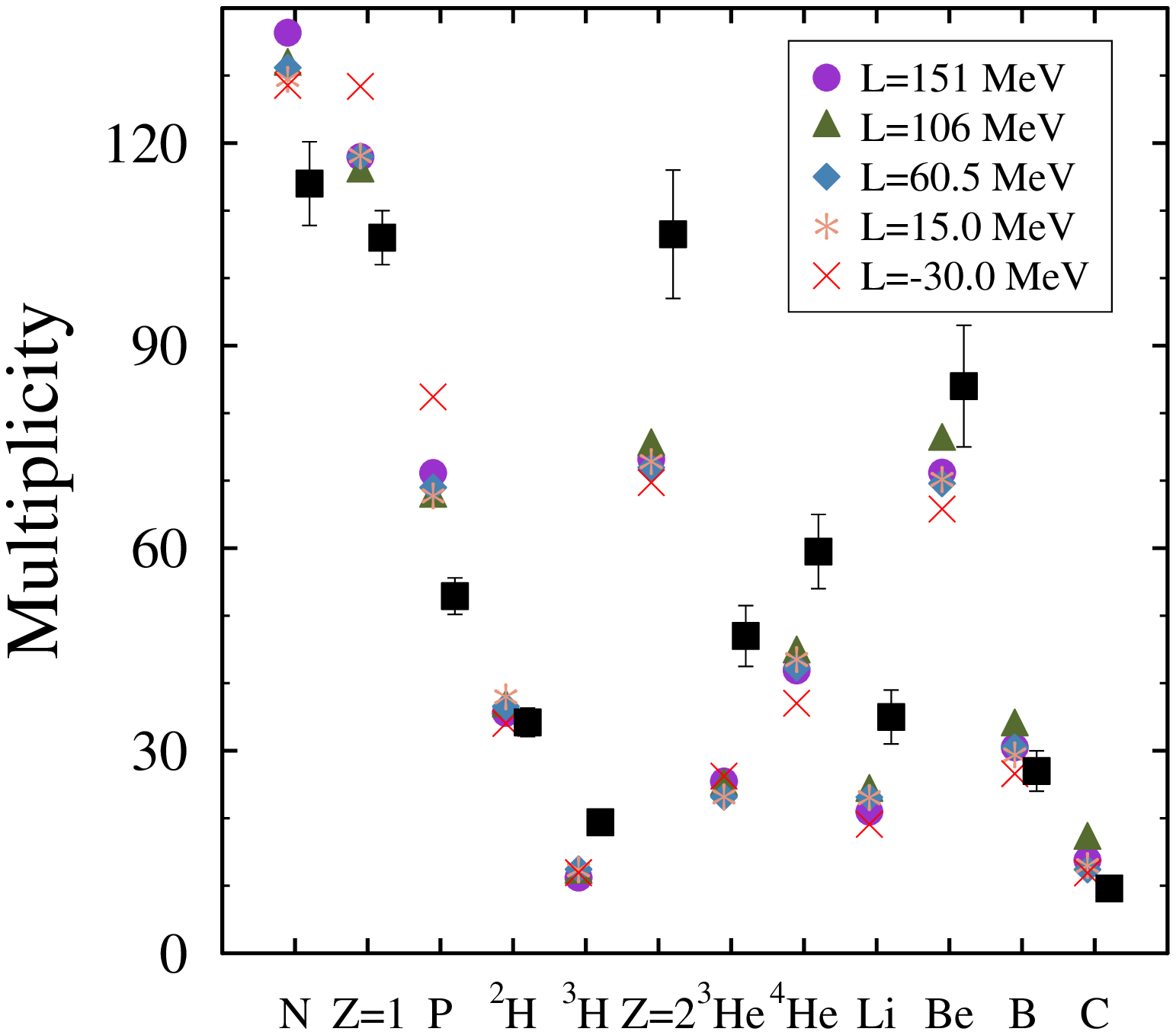}}
\end{center}
%\end{minipage}
\caption{(Left panel) Multiplicities of free nucleons and clusters with $Z~\leq$~6 for three values of the 
momentum space coalescence parameter ($\delta p$=0.15, 0.20 and 0.25 GeV/c) compared to 
the experimental FOPI data~\cite{Reisdorf:2010aa} (full squares with error bars). 
The asy-EoS stiffness parameters have been set to $L$=80.0 MeV and $K_{sym}$=0.0 MeV.
The values of the coalescence parameters are listed in \tabref{rcoalpars}.
(Right panel) Dependence of free nucleon and cluster multiplicities on the stiffness of the symmetry
energy. The $p$-space coalescence parameter has been set to $\delta p$=0.20 GeV/c while the $r$-space
coalescence parameters have been determined in each case from a fit to the shown experimental data. 
Multiplicities have been multiplied, for each specie, by the indicated scaling factor shown in the left panel.
 }
\figlab{clustmult}
\end{figure*}

The values of the five coalescence parameters are determined from a fit of cluster and free nucleon
multiplicities to experimental data. For the present case of Au+Au collisions at an impact energy
of 400 MeV/nucleon the FOPI cluster multiplicity experimental data for central collisions ($b\leq$ 2.0 fm)
have been used ~\cite{Reisdorf:2010aa}, namely the multiplicities of $p$, $n$, $^2$H, $^3$H, 
$^3$He, $^4$He, Li, Be and C have been fitted. The neutron multiplicity has not been measured
experimentally. Its value has been deduced by subtracting the number of neutrons bound in experimentally
detected clusters from the total of 236. The quality of the fit is impacted only marginally by the incorporation
of this information in the multiplicity fit. 

A correlation between the values of the $r$ and $p$-space coalescence parameters has been observed.
Lower values for the latter lead to higher values for the former. Acceptable values for the two
should be of the order of the range of nuclear forces and close to the Fermi momentum of nucleons in nuclei.
%The extracted values of coalescence parameters depend on the chosen equation of state, in particular
%the symmetry energy stiffness, and the coalescence time $t_C$. Softer choices for the former and larger
%values for the latter lead to larger values for the $r$-space coalescence parameters. It has been
%noted that smaller ($\approx$0.1 GeV/c)  values for the momentum space coalescence parameters 
%$\delta p_n$ and $\delta p_p$ allow a better description of experimental cluster multiplicities,
%see left panle of \figref{clustmult}. This is achieved at the expence of larger values for the $r$-space coalescence
%parameters, which should however be comparable (or smaller) to the range of the nuclear force. 
It has also been noticed that the quality of the fit is not altered if the constraint 
$\delta p_p$=$\delta p_n$=$\delta p$ is enforced. 

By performing tests with $\delta p$ in the range 0.1-0.3 GeV/c
the middle-ground value $\delta p$=0.2 GeV/c has been selected for the results presented in this paper.
It allows a compromise between the best possible description of experimental cluster and free nucleons multiplicities 
and values of $r$-space coalescence parameters comparable to the range of the strong force for coalescence
times ($t_C$) in the range 100-150 fm/c. For values $t_C$$\geq$125 fm/c the independence of theoretical elliptic
flow ratios values on this parameter is also achieved. Consequently all results reported in this paper have been derived 
by setting $t_C$=150 fm/c. It is worth mentioning that the impact of varying $\delta p$ in the specified range 
on the extracted values of $L$ and $K_{sym}$ is significantly smaller than the uncertainty
induced by the inaccuracy of presently available experimental data for elliptic flow ratios.

A comparison between experimental multiplicities for clusters with $Z$$\leq$6 in central
($b\leq$2 fm) $^{197}$Au+$^{197}$Au collisions at 400 MeV/nucleon impact energy and theoretical results of
the fitted model for three different values of the $p$-space coalescence parameter $\delta p$=0.15, 
0.20 and 0.25 GeV/c is presented in the left panel of \figref{clustmult}. It is readily observed that multiplicities of neutrons,
protons and helium isotopes depend rather strongly on the value of this parameter, the opposite being true for
deuterons and tritons. Also, for all Z$\leq$2 isotopes smaller values lead to closer agreement to experimental
data. In fact, for even smaller values ($\delta p$=0.10 GeV/c), not shown here, the model is able to reproduce
experimental data at 2$\sigma$ confidence level, at the expense of unrealistically large values for the $\delta r$ parameters.
%This may suggest that a better
%description of experimental cluster multiplicity data may be possible, for reasonable values of the coalescence
%parameters, for QMD-like transport models that feature an improved stability of the initialized nuclei.

Another noteworthy feature of the results presented in~\figref{clustmult} is that the experimental deuteron multiplicity 
is closely reproduced, in opposition to similar models reported in the literature \cite{Russotto:2011hq}.
The same holds true for Be, B and C nuclei. The multiplicities of Li, He isotopes and triton are
under-predicted, but the discrepancy to experimental data is smaller as compared to results reported elsewhere.
The reason behind this improvement has been identified to be the momentum dependence of the MDI2 interaction.
In contrast, the MDI interaction leads to deuteron multiplicities that under-predict the experimental values by
at 30-50\%, depending on the stiffness of the symmetry energy.

The right panel of ~\figref{clustmult} displays the dependence of free nucleon and cluster multiplicities on
the stiffness of the symmetry energy for five values of the slope parameter $L$. For each case the value
of the curvature $K_{sym}$ corresponds to the one shown in the upper half part of~\tabref{lsymxyvals}. 
While the $p$-space coalescence parameter has been kept fixed to $\delta p$=0.20 GeV/c, the $r$-space
coalescence parameters have been determined in each case from fits to the FOPI experimental data.
Such an approach is legitimate since the stiffness of the symmetry energy is not a priori known. The quality
of the fit is however not significantly affected if the $r$-space coalescence parameters were considered as independent
of the SE stiffness, taking them, for example, as the average of the values obtained from the stiffness dependent fits.
The dependence of multiplicities on the stiffness of the asy-EoS is not monotonic and for certain cluster
species is negligible ($^3$H and $^3$He) suggesting that additional physics input is needed in the transport
model in order to allow a precise description. The sensitivity of cluster multiplicities to the stiffness of the
symmetry energy is of similar magnitude as to the value of the $\delta p$ parameter. 

\begin{figure*}
\begin{center}
\resizebox{0.45\textwidth}{!}{
\includegraphics{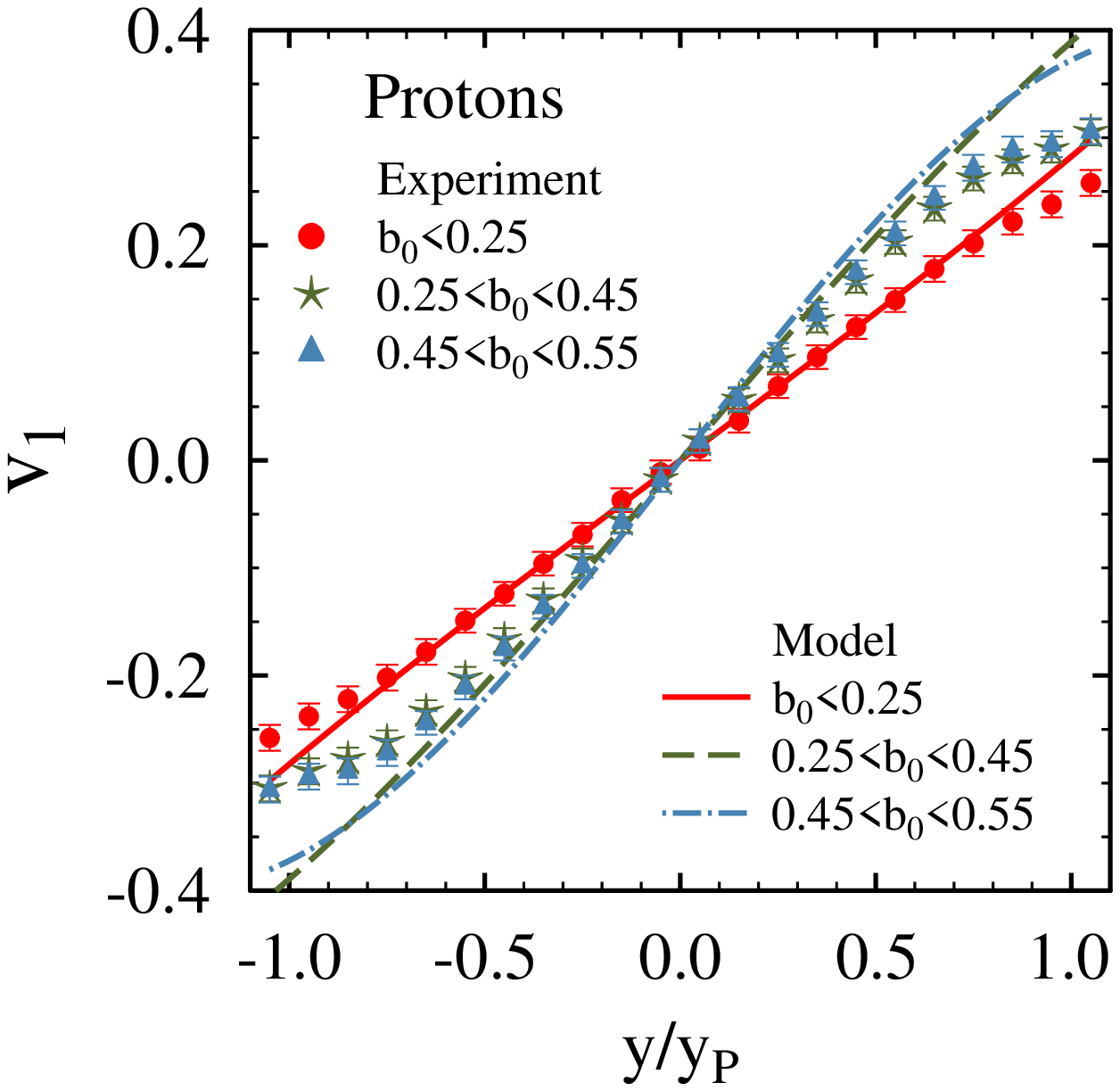}}
\resizebox{0.45\textwidth}{!}{
\includegraphics{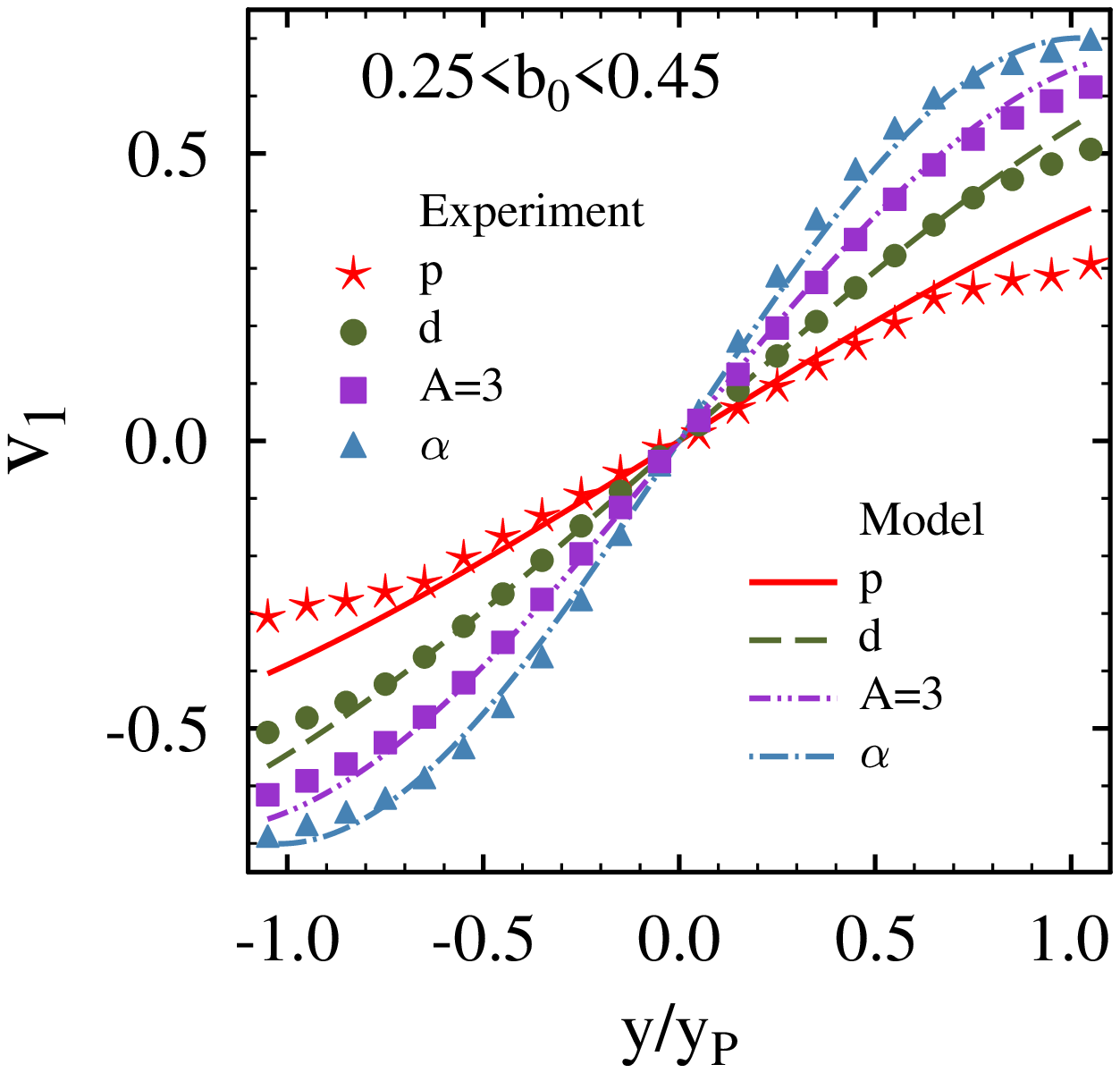}}
\end{center}
\caption{(Left panel) Transverse flow of protons as a function of center-of-mass reduced rapidity ($y_P$ stands for
the projectile's rapidity) for three impact parameter ranges. (Right panel) Transverse flow of protons and
light clusters in mid-central collisions as a function of center-of-mass relative rapidity. In both cases
the corresponding FOPI experimental data~\cite{FOPI:2011aa} for $^{197}$Au+$^{197}$Au at an impact energy
of 400 MeV/nucleon are shown for comparison. The kinematical cut $p_T/p_P>$ 0.4 has been applied to results,
with $p_P$ the projectile's momentum in the center-of-mass frame. The reduced parameter
$b_0$=$b/b_{max}$ with $b_{max}$=13.4 fm has been used to express centrality ranges.}
\figlab{v1fopi}
\end{figure*}

\begin{figure*}
\begin{center}
\resizebox{0.45\textwidth}{!}{
\includegraphics{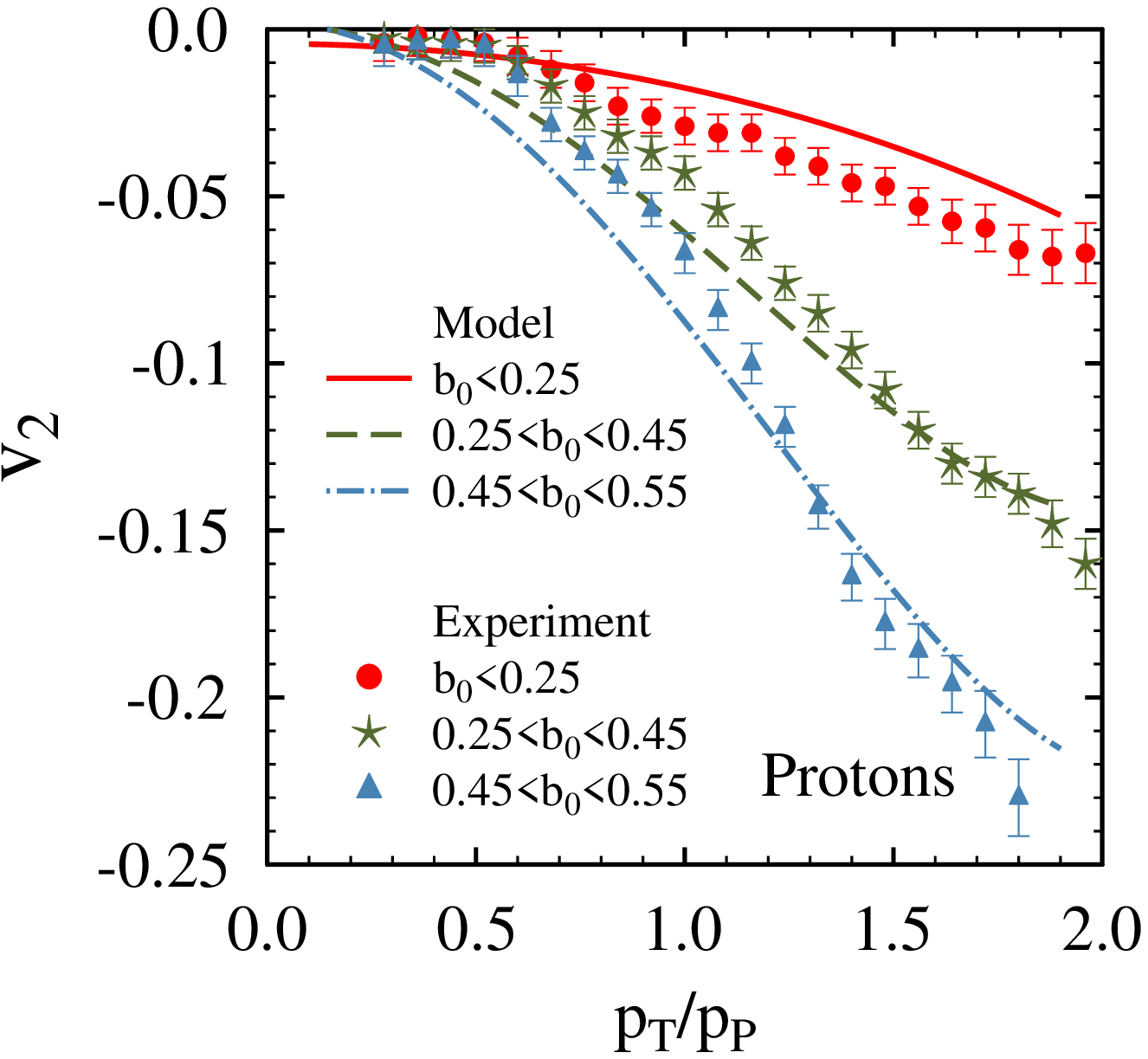}}
\resizebox{0.45\textwidth}{!}{
\includegraphics{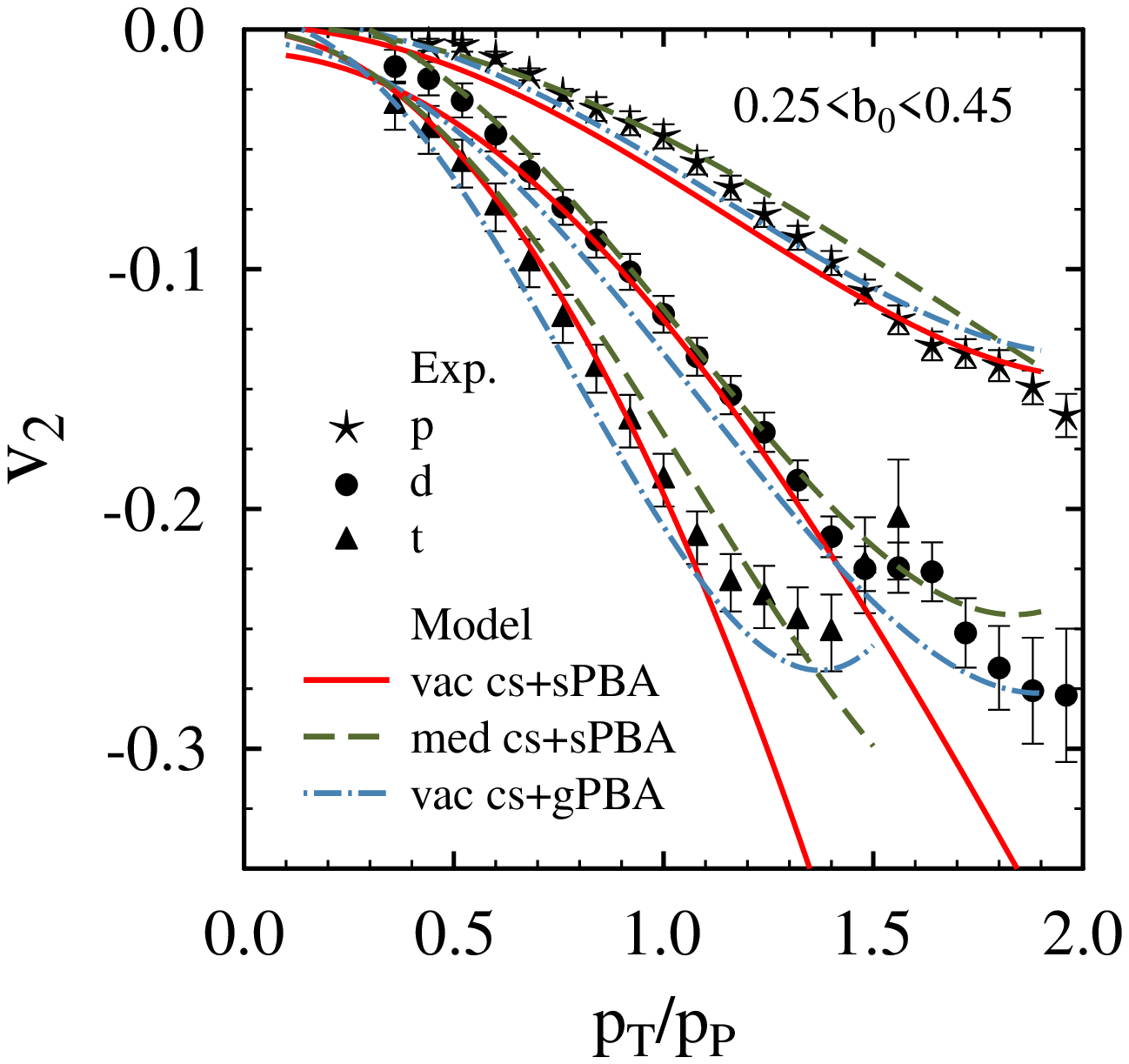}}
\end{center}
\caption{(Left panel) Elliptic flow of protons as a function of relative transverse
momentum $p_T/p_P$ for three impact parameter ranges; (Right panel) Elliptic flow of protons, 
deuterons and tritons in mid-central
collisions as a function of the relative transverse momentum. For each particle species calculations with
the indicated choices for elastic cross-sections, vacuum (``vac cs'') or in-medium (``med cs''), and
Pauli blocking algorithm (sPBA or gPBA) are shown, see text for details. In each case the kinematical cut $|y/y_P|<$0.4
has been applied. The same observation for the experimental data set and reduced impact parameter $b_0$ as for
~\figref{v1fopi} are in order.}
\figlab{v2fopi}
\end{figure*}

\section{Model validation}
\seclab{modelvalid}
\subsection{Transverse and Elliptic Flows}
\seclab{thexpv2comp}
Predictions of the model described in the previous section have been compared with relevant
available experimental data to test its validity. A comparison with transverse and elliptic flow
data due to the FOPI~\cite{FOPI:2011aa}, FOPI-LAND~\cite{Cozma:2013sja} and ASYEOS~\cite{Russotto:2016ucm}
collaborations will be presented in this section.

The following setting of model parameter has been used, if not otherwise stated: the MDI2 potential
has been used with the stiffness parameters of the isoscalar EoS as mentioned in~\tabref{model_input_params},
the stiffness parameter of the asy-EoS have been set to $L$=80.0 MeV and $K_{sym}$=0.0 MeV (close to the central
values of these parameters extracted in this work, see~\secref{lksymmdi2}), vacuum parametrizations for elastic
$NN$ scattering cross-sections and sPBA Pauli blocking algorithm.

Model predictions for transverse and elliptic flows compared to experimental FOPI data are shown in
\figref{v1fopi} and \figref{v2fopi} respectively. The left panel of each figure presents proton flows
for three centrality ranges $b_0$$<$0.25, 0.25$<$$b_0$$<$0.45 and 0.45$<$$b_0$$<$0.55 where $b_0$ is the reduced
impact parameter. Model predictions generally agree reasonably well with experimental values, some deviations are
however observed in mid-peripheral collisions for the transverse flow $v_1$ and mid-central collisions
for the elliptic flow $v_2$ for lower values of the reduced transverse momentum. An improved description of $v_1$ may be obtained by a finer tuning
of the compressibility modulus $K_0$ and skewness parameter $J_0$ as the transverse flow is sensitive
to them~\cite{Danielewicz:2002pu}. This is however outside the scope of the present study.

The right panels of \figref{v1fopi} and \figref{v2fopi} present elliptic flows of protons and light clusters
in mid-central collisions compared to published FOPI data. Model predictions for transverse flow 
of light clusters (deuteron, $A$=3 and $\alpha$ particle) describe experimental values visibly better
than in the case of protons. For the case of elliptic flow the model generally predicts a stronger
effect than measured experimentally for every particle species (protons, deuterons, tritons). 
This observable is however affected by important uncertainties
due to in-medium effects on elastic cross-sections and the Pauli blocking algorithm used. By using either
empirical FU3FP4 in-medium cross-sections parametrizations~\cite{Wang:2013wca,Li:2011zzp} or the gPBA Pauli blocking algorithm as opposed
to the standard choices a better description of the FOPI experimental values for $v_2$ can be achieved.
These modifications change the predictions for $v_1$ only very slightly.

A comparison of model predictions and experimental FOPI-LAND data is shown in \figref{v2fopiland}. Predictions for elliptic
flows of neutrons, protons and hydrogen for different stiffnesses of the asy-EoS are presented. It is seen
that while theoretical predictions for the standard choice of model parameters do agree with the experimental
values of flow for a certain value of the stiffness parameter, this value of the stiffness parameter
depends strongly on the particle species. This can be traced back to a rather strong dependence of $v_2$
values on the compressibility modulus of symmetric nuclear matter, in-medium effect on elastic cross-sections and the Pauli
blocking algorithm used. This sensitivity is illustrated by presenting model predictions for three values
of the compressibility modulus, $K_0$=210, 245 and 280 MeV and by switching between vacuum and empirical
in-medium elastic cross-sections~\cite{Wang:2013wca,Li:2011zzp} parametrizations or between sPBA and gPBA Pauli
blocking algorithms. The impact on the magnitude of $v_2$ can amount to as much as 40$\%$ of the value
obtained using the standard choice of model parameters.

In \figref{v2asyeos} a comparison between model predictions and the recent experimental data of
the ASYEOS collaboration~\cite{Russotto:2016ucm} for elliptic flow of neutrons and charged particles is shown.
Only calculations for the standard choice of model parameters ($K_0$=245 MeV, vacuum cross-sections and sPBA
Pauli blocking algorithm) are presented. As before, the stiffness of the SE has been adjusted by changing the
value of the $x_{MDI}$ parameter. Elliptic flow of neutrons is more sensitive to the density dependence of the SE,
as compared to elliptic flow of clusters. 

By comparing \figref{v2asyeos} and \figref{v2fopiland} its is evident that the description of experimental
data elliptic flow of neutrons due to the ASYEOS and FOPI-LAND collaborations is of similar quality,
requiring a similar value for $K_0$ for a proper description. Surprisingly this is not the case for the
elliptic flow of hydrogen (FOPI-LAND) and charged particles (ASYEOS). The latter would require a sensibly
stiffer compressibility modulus than the standard $K_0$=245.

\begin{figure}
\begin{center}
\resizebox{0.45\textwidth}{!}{
\includegraphics{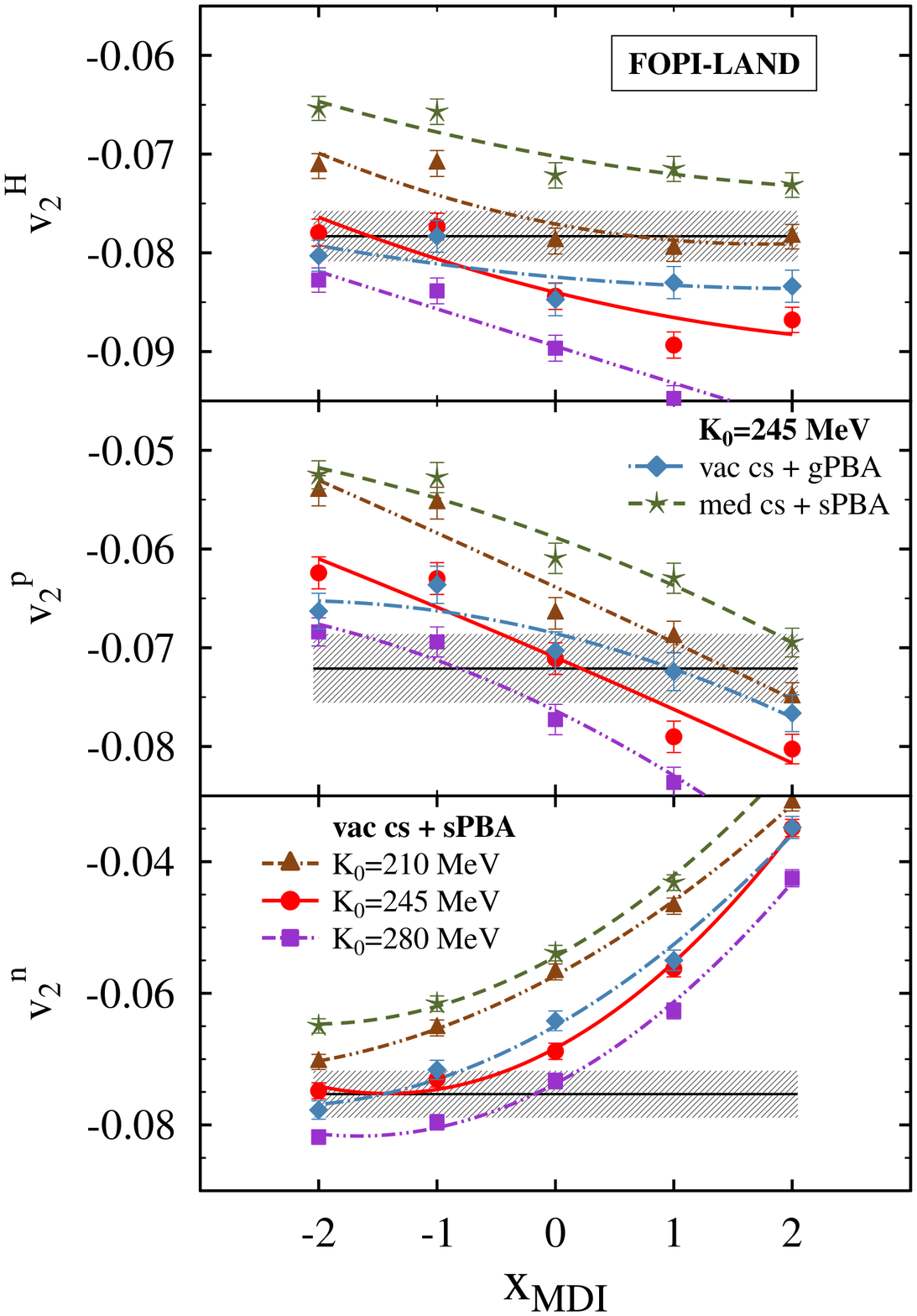}}
\end{center}
\caption{Theoretical predictions for elliptic flow of neutrons (bottom), protons (middle)
and hydrogen (top) as a function of the stiffness parameter $x_{MDI}$ (see~\tabref{lsymxyvals} for
the corresponding values of $L$ and $K_{sym}$). Calculations using the indicated combinations for
elastic cross-sections and Pauli blocking algorithms are shown. For the case ``vac cs+sPBA'' calculations
for three values of the compressibility modulus $K_0$ have been performed, while for the rest only the standard
$K_0$=245 MeV has been chosen. The FOPI-LAND set B integrated impact parameter~\cite{Cozma:2013sja} data,
multiplied by a factor 1.15 to include the omitted reaction plane dispersion correction 
factor~\cite{Trautmann:2013aa}, are shown for comparison. Theoretical spectra have been filtered accordingly.}
\figlab{v2fopiland}
\end{figure}

\begin{figure}
\begin{center}
\resizebox{0.45\textwidth}{!}{
\includegraphics{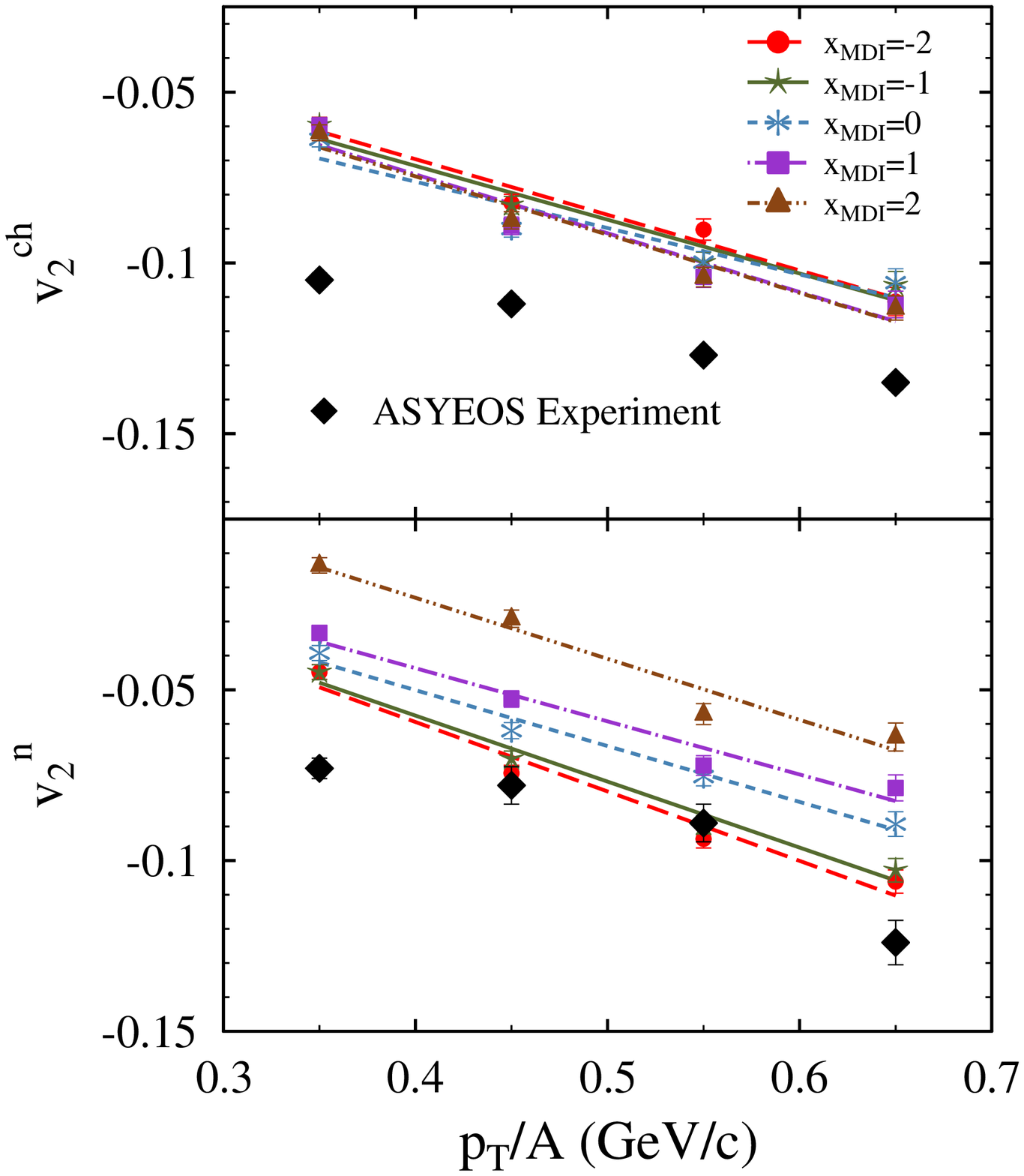}}
\end{center}
\caption{Theoretical predictions for elliptic flow of neutrons (bottom) and charged particles (top)
as a function of the transverse momentum per nucleon compared to the ASYEOS data~\cite{Russotto:2016ucm}.
Results for different values of the stiffness parameter $x_{MDI}$ are shown. Calculations have been
performed with the standard choice of model parameters and have been subjected to the ASYEOS filter (both 
kinematics and efficiency of particle detection).}
\figlab{v2asyeos}
\end{figure}

An obvious source of systematic uncertainties
for model predictions of hydrogen and charged particles flows is the inability of coalescence models to
reproduce the experimental values of proton-to-clusters multiplicity ratios. To partially account for this
effect theoretical corrected values for elliptic flow of hydrogen ($\tilde v_2^H$) and charged
clusters ($\tilde v_2^{ch}$) can be determined by employing the theoretical
value of elliptic flow and the experimental multiplicity for each cluster species
\begin{eqnarray}
\eqlab{v2corr}
 \tilde v_2^H&=&\frac{M_p^{exp}\,v_2^p+M_d^{exp}\,v_2^d+M_t^{exp}\,v_2^t}{M_p^{exp}+M_d^{exp}+M_t^{exp}} \\
 \tilde v_2^{ch}&=&\frac{M_p^{exp}\,v_2^p+\sum_{Z_i \ge 1,N_i \ge 1} M_{Z_i,N_i}^{exp}\,v_2^{Z_i,N_i}}
 {M_p^{exp}+\sum_{Z_i \ge 1,N_i \ge 1} M_{Z_i,N_i}^{exp}}\,. \nonumber
\end{eqnarray}
This approach is justifiable since theoretical elliptic flows of individual cluster species 
can describe experimental data for mid-central collisions reasonably well, see~\figref{v2fopi}.
In the above expressions the available experimental multiplicities of clusters in central collisions
~\cite{Reisdorf:2010aa} have been used as a first approximation, since values for the 
needed $b<$7.5 fm case are not available in the literature.
The correction factors $f_{corr}^H=\tilde v_2^H/ v_2^H$ and $f_{corr}^{ch}=\tilde v_2^{ch}/ v_2^{ch}$
depend on the various model parameters, in particular asy-EoS stiffness. Their average values were determined to be
$f_{corr}^H$=1.075 and $f_{corr}^{ch}$=1.10. The exact values, determined using the appropriate multiplicities,
may be somewhat higher, particularly for $f_{corr}^{ch}$, since multiplicities of intermediate mass
fragments increase with impact parameter up to $b$=8.0 fm \cite{Schuttauf:1996ci}. Similar correction
factors can be determined for the case $b<$ 2.0 fm, obtaining the moderately higher values
$f_{corr}^H$=1.11 and $f_{corr}^{ch}$=1.14. Conservative ranges for these parameters may thus be
given by $f_{corr}^H$=1.075$\pm$0.05 and $f_{corr}^{ch}$=1.10$\pm$0.05.
A potentially superior approach would be to determine, both theoretically
and experimentally, coalescence invariant elliptic flows of neutrons and protons, similar to the
case of coalescence invariant neutron and proton multiplicity spectra~\cite{Famiano:2006rb} or develop
transport models that account directly for light cluster degrees of freedom~\cite{Danielewicz:1991dh}.

Multiplying theoretical $v_2^{ch}$ values for the ASYEOS case in~\figref{v2asyeos} by $f_{corr}^{ch}$
shifts predictions toward the experimental data, a discrepancy of about $20-25\%$ persists
for lower $p_T$ values.
%An improvement could be obtained by increasing the compressibility modulus 
%of symmetric nuclear matter to values that are however unrealistically large.
For the case of the FOPI-LAND data a scaling of $v_2^H$ with $f_{corr}^H$ makes a stronger case
for $K_0$=210 MeV, see~\figref{v2fopiland}.

\section{Density Dependence of Symmetry Energy}
\seclab{ddse}
\subsection{Previous Version of the Model}
\seclab{prevmodel}
A previous version of the model~\cite{Cozma:2011nr,Cozma:2013sja} has been used to study the density dependence
of the symmetry energy using elliptic flow observables. Constraints for the slope of the symmetry energy have been extracted
from the FOPI-LAND experimental data for neutron-to-proton EFR and EFD: L=118$^{+45}_{-57}$ MeV (npEFR)
and L=129$^{+46}_{-80}$ MeV (npEFD) respectively~\cite{Cozma:2013sja}. As already noted in that study the central values
are significantly larger than the ones extracted from an analysis of neutron skin thickness and isospin diffusion
at lower energies, L$\approx$70 MeV~\cite{Tsang:2012se}. A smaller difference
with respect to similar analyses of the transverse momentum dependent version of the FOPI-LAND 
data using the UrQMD transport model~\cite{Russotto:2011hq,Wang:2014rva}, which arrive at the 
constraint $L$=89$\pm$45 MeV (at 2$\sigma$ CL), has been found. In these last cases part of the
discrepancy could be understood as originating from the different methods that where used to
analyze experimental data: ERAT + multiple centrality bins~\cite{Cozma:2013sja} 
and multiplicity + one centrality bin~\cite{Russotto:2011hq,Wang:2014rva}. In this context it is worth
mentioning that the more precise value of the slope $L$ extracted from the recent ASYEOS collaboration
data~\cite{Russotto:2016ucm} is in full agreement with the previous studies of the same group~\cite{Russotto:2011hq}.

\begin{figure}
\begin{center}
\resizebox{0.45\textwidth}{!}{
\includegraphics{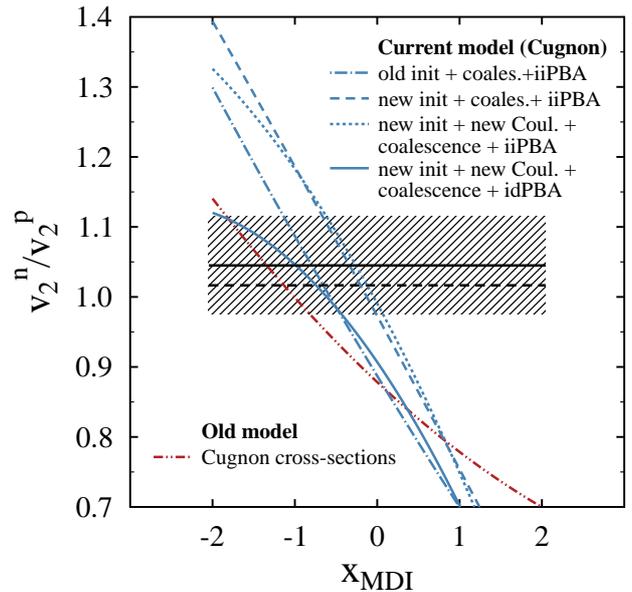}}
\end{center}
\caption{Predictions for the npEFR for the previous~\cite{Cozma:2013sja}
and current versions of the transport model. The differences between the two are
added incrementally, making clear their relative importance and the impact on the
extracted constraints for the symmetry energy stiffness, see text for explanations.
The MDI potential and the Cugnon parametrization of cross-sections have been used in the calculations.
The values of $L$ and $K_{sym}$ corresponding to the shown integer values of $x_{MDI}$ are given
in \tabref{lsymxyvals}. Filtered theoretical results are compared with FOPI-LAND experimental data.
Experimental results using ERAT + multiple centrality bins (full line) and multiplicity + 1 centrality bin (dashed line)
for impact parameter determination are depicted by horizontal lines. For the former case the uncertainty is depicted by a 
hashed horizontal band.
}
\figlab{oldvsnewmodel}
\end{figure}

The study in Ref.~\cite{Cozma:2013sja} has made use of a density cut-off algorithm to determine final spectra
of neutrons and protons, which were considered as free if they were located in a region of density less than $\rho_0$/8
at clusterization time. Moreover, the value of the cut-off density has not been adjusted to reproduce as closely
as possible measured free nucleon multiplicities. To make use of existing data for EFR of neutron-to-hydrogen
and neutron-to-charged particles the coalescence model described in~\secref{coalescence} has been developed. As described in~\secref{ininuc}
the initialization part of the model has been improved to describe nuclear density profiles more accurately.
In the context of the study of pion production in heavy-ion collisions~\cite{Cozma:2016qej} the strength of the
Coulomb interaction has been slightly modified to fit its contribution to binding energy as determined from
nuclear mass formulae. Additionally, the isospin independent Pauli blocking algorithm employed previously has
been replaced by an isospin dependent one.

The impact of these modifications on the value of the npEFR and the extracted constraint for the SE
stiffness is presented in \figref{oldvsnewmodel}. For this calculation the old MDI potential together with the
Cugnon parametrization of cross-sections and a value of the compressibility modulus $K_0$=210 MeV have been employed. 
Starting from the old calculation (dashed-double dotted
curve) the model modifications described above are switched on incrementally, first the coalescence algorithm
(dash-dotted curve) followed by the improved initialization of nuclei (dashed curve), the modified
Coulomb interaction (dotted curve) and finally the isospin dependence in the Pauli blocking algorithm (full curve).
The before last modification has little impact on EFR and the extracted value for the
slope $L$. The other three improvements have a comparable and non-negligible impact, the first two leading to a softer
constraint for the SE stiffness while the last one leads to a stiffer asy-EoS. The combined effect of these model modifications
leads to the new constraints:  $L$=108$^{+36}_{-27}$ MeV (npEFR) and $L$=91$^{+39}_{-24}$ (npEFD).
Similar conclusions hold for neutron-to-hydrogen EFR and EFD. The difference with respect to the results
presented in \secref{lksymcmdi2} is mostly due to the softer compressibility modulus, the impact of the different parametrization
used for elastic cross-sections and modification of the optical potential at high momenta are of secondary importance.

It is worth noting that values of the same observables determined using different methods to analyze experimental data
lead to constraints for the value of $L$ that differ by 12 MeV (softer) if neutron-to-proton observables are used. 
It has been noted previously~\cite{Wang:2014rva} that the impact is even larger, of the order 25-35 MeV, if neutron-to-hydrogen
observables are used.

\begin{figure*}
\begin{center}
\resizebox{0.45\textwidth}{!}{
\includegraphics{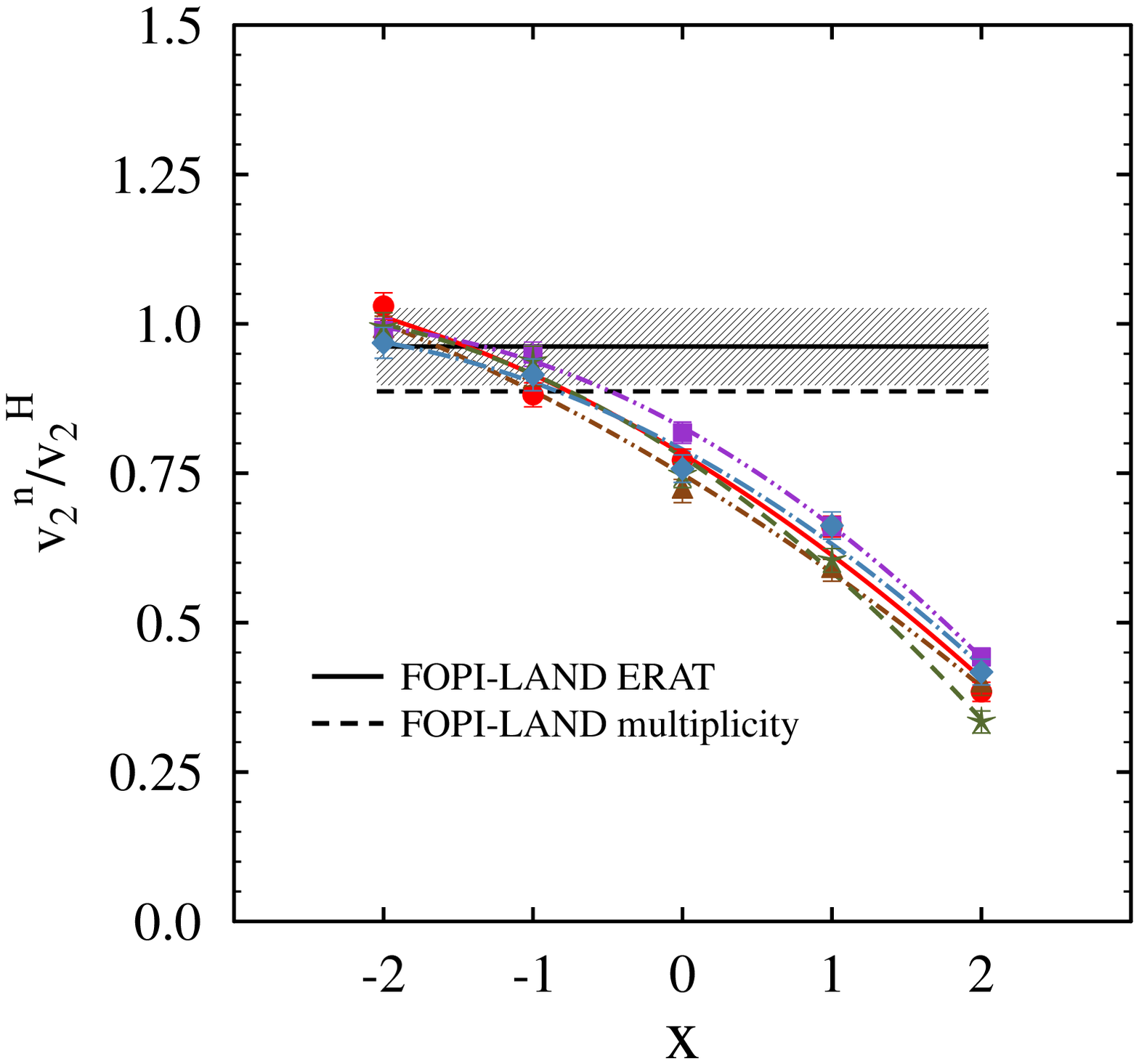}}
\resizebox{0.45\textwidth}{!}{
\includegraphics{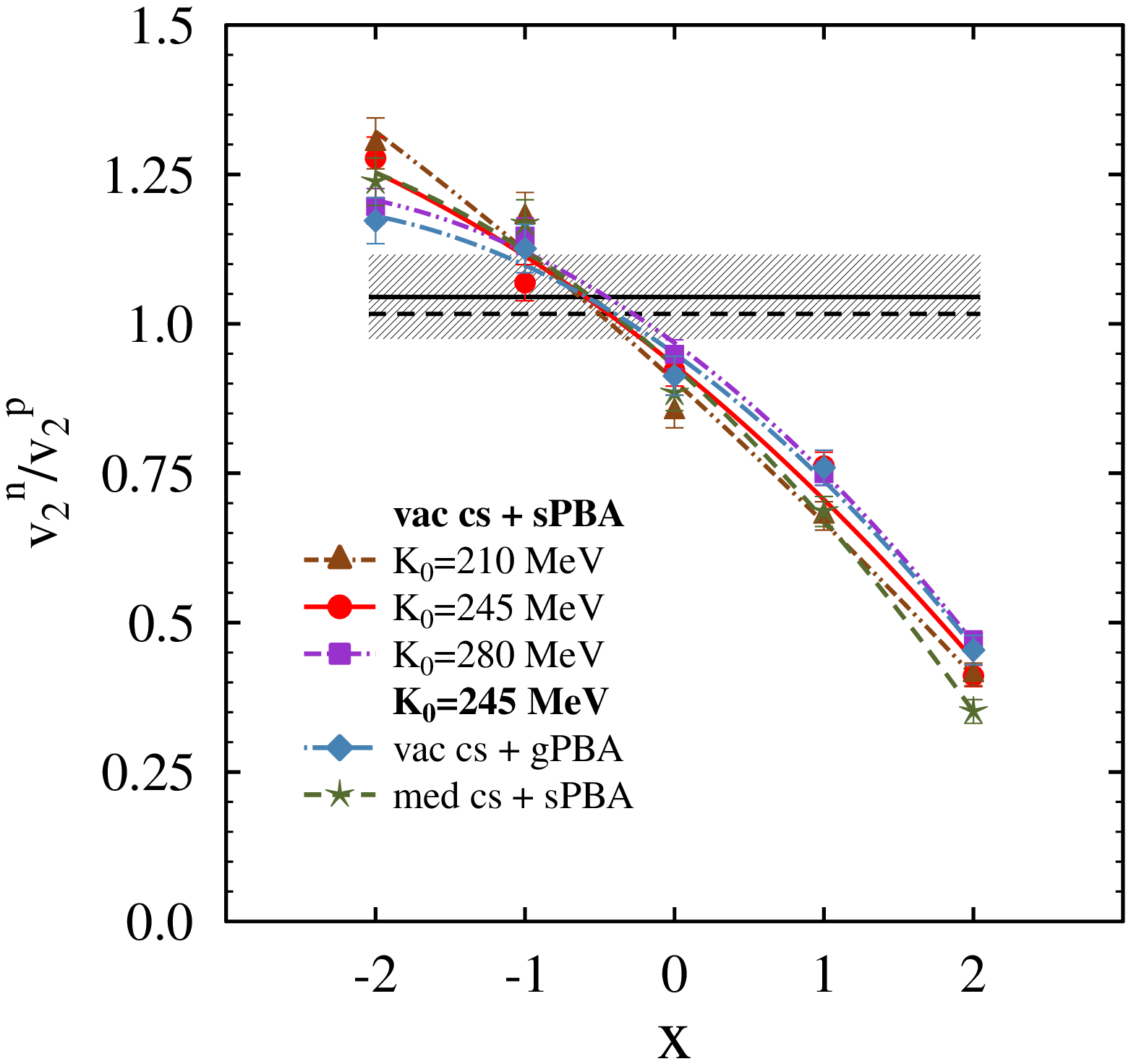}}
\end{center}
\caption{Neutron-to-hydrogen (left panel) and neutron-to-proton (right panel) EFR as
a function of the asy-EoS stiffness parameter $x_{MDI}$ are compared to FOPI-LAND data. 
The same comments as in \figref{v2fopiland} apply to theoretical predictions. Experimental
values that where obtained using ERAT + multiple centrality bins~\cite{Cozma:2013sja} and 
multiplicity + 1 centrality bin~\cite{Wang:2014rva} for impact parameter determination are shown
by horizontal lines. Only the uncertainty corresponding to
the former data set is shown (horizontal hashed band).}
\figlab{efrfopiland}
\end{figure*}

\subsection{Constrained MDI2 potential}
\seclab{lksymcmdi2}
It is customary to extract the stiffness of the symmetry energy using parametrizations that employ a single
free parameter for that purpose. The MDI~\cite{Das:2002fr} potential belongs to that category,
a parameter denoted $x$ had been introduced to allow changes of the asy-EoS stiffness. As
already pointed out, this leads to a constraint between $L$ and $K_{sym}$ and the two-dimensional 
parameter space associated to these quantities is reduced to a one-dimensional one (curve). 
Integer values of $x$ have been used to simulate heavy-ion reactions with different 
stiffnesses for the SE. The corresponding values for $L$ and $K_{sym}$ are shown in~\tabref{lsymxyvals}
together with the values of $x$ in MDI, denoted as $x_{MDI}$ to distinguish it from the parameter $x$
appearing in the MDI2 potential. By adjusting the values of the parameters $x$ and $y$ appearing in MDI2
to the values listed in~\tabref{lsymxyvals} a potential with the same $L$-$K_{sym}$ combinations as for MDI,
labeled cMDI2, can be obtained. The momentum dependence of cMDI2 and MDI potentials is however different.

The potential cMDI2 has been employed to simulate $^{197}$Au+$^{197}$Au collisions with an impact energy
of 400 MeV per nucleon. Theoretical results for elliptic flow of neutrons, protons, hydrogen and charged particles
have been compared to the corresponding experimental results obtained by the FOPI-LAND~\cite{Cozma:2013sja}
and ASYEOS~\cite{Russotto:2016ucm} collaborations in \secref{thexpv2comp}. A comparison
of the theoretical and experimental elliptic flow ratios relevant for the extraction of 
the SE stiffness are presented in~\figref{efrfopiland} and \figref{efrasyeos} for the
FOPI-LAND and ASYEOS cases respectively. The extracted values for the slope parameter $L$
using all experimentally measured observables and certain combinations of them
are presented in~\tabref{lsym_mdi_diffobs}.

Firstly, it is noted that the extracted value of $L$ is impacted by different methods employed
to analyze the experimental FOPI-LAND data. 
Analyzing experimental data using multiple centrality bins and the ERAT observable
to determine impact parameter and alternatively one centrality bin and multiplicities 
lead to slope parameter values that differ by $\delta L$=39 MeV if the nhEFR is used.
For npEFR the discrepancy is reduced to $\delta L$=9 MeV, well below the quoted uncertainty
due to other systematical and statistical error effects. The culprit was found to be primarily
related to the number of centrality bins used, and only marginally due to the observable used to extract
the value of the impact parameter~\cite{Trautmann:2013aa}.

Secondly, constraints for $L$ extracted using ASYEOS data are systematically below those obtained
using FOPI-LAND data. For the integrated nchEFR the difference with respect to npEFR result
is even slightly outside the 1$\sigma$ CL compatibility region. This discrepancy can be understood if
theoretical EFR are computed by using the corrected values $\tilde v_2^H$ and $\tilde v_2^{ch}$,
introduced in~\eqref{v2corr}, for elliptic flow of hydrogen and charged particles respectively.
The new extracted constraints for $L$, labeled as $L_{corr}$ in the rightmost column of
\tabref{lsym_mdi_diffobs}, are now in better agreement with each other. However the extracted
value for $L_{corr}$ using nhEFR is now barely compatible with that corresponding to npEFR.
Systematical uncertainties affecting theoretical values for hydrogen
and charged particles flows due to under-predicted values for proton-to-cluster multiplicity ratios
will thus have to be better understood and eliminated before a fully consistent picture for the extracted
stiffness of SE using npEFR, nhEFR and nchEFR observables can be achieved and the available
experimental data sets can be used to their full potential.

Lastly, values for the extracted values of $L$ and $L_{corr}$ using certain combinations of observables
are presented at the bottom of~\tabref{lsym_mdi_diffobs}. Since npEFR and nchEFR probe, on average,
different densities, npEFR + nchEFR and alternatively npEFR + $p_T$ dependent nchEFR can be used to obtain
independent constraints for both $L$ and $K_{sym}$. The value of the slope $L$, together with
the corresponding $K_{sym}$, extracted using the cMDI2 potential will be used in \secref{lksymmdi2} 
as a benchmark for those results. The same observations as above apply when comparing
to results obtained using different experimental data sets or the impact of systematic
uncertainties of theoretical cluster multiplicities.

\begin{figure}
\begin{center}
\resizebox{0.45\textwidth}{!}{
\includegraphics{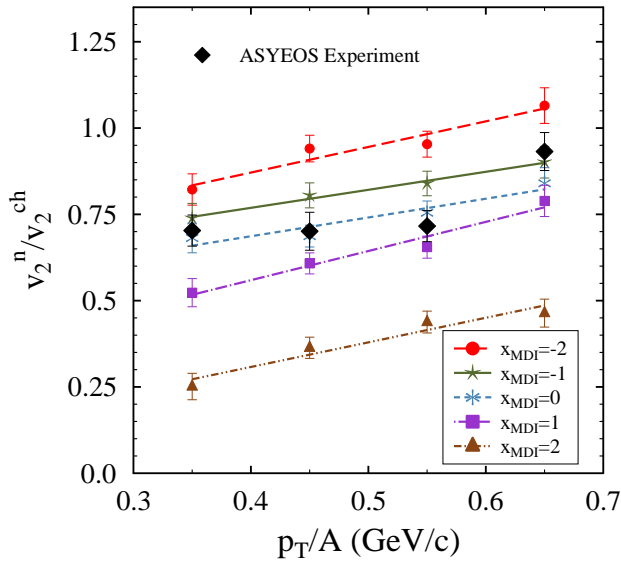}}
\end{center}
\caption{Theoretical $p_T$ dependent neutron-to-charged particles elliptic flow ratios are compared to
experimental ASYEOS data~\cite{Russotto:2016ucm}. Results for different stiffnesses of the symmetry energy
are shown.}
\figlab{efrasyeos}
\end{figure}

\begin{table}
\caption{Constraints for the slope parameter $L$ and its corrected value $L_{corr}$ 
extracted using available FOPI-LAND and ASYEOS experimental data. Model parameters
have been set to their standard values.}
\tablab{lsym_mdi_diffobs}       % Give a unique label
\begin{center}
\begin{tabular}{c|c|c}
\hline\noalign{\smallskip}
Observable & L [MeV] & L$_{\mathrm{corr}}$ [MeV]\\
\noalign{\smallskip}\hline\noalign{\smallskip}
FOPI-LAND (ERAT) & & \\
n/p  & 89$^{+24}_{-22}$ & 89$^{+24}_{-22}$\\
n/H  & 133$^{+25}_{-29}$ & 163$^{+17}_{-21}$\\
\noalign{\smallskip}\hline\noalign{\smallskip}
FOPI-LAND (mult) &  &\\
n/p  & 80$^{+26}_{-23}$ & 80$^{+26}_{-23}$\\
n/H  & 94$^{+29}_{-24}$ & 128$^{+27}_{-29}$\\
n/H (p$_T$) & 85$^{+51}_{-51}$ & 132$^{+60}_{-60}$\\
\noalign{\smallskip}\hline\noalign{\smallskip}
ASYEOS & &\\
n/ch & 39$^{+6}_{-6}$ & 63$^{+8}_{-8}$\\
n/ch (p$_T$) & 65$^{+25}_{-25}$ & 95$^{+26}_{-26}$\\
\noalign{\smallskip}\hline\noalign{\smallskip}
FOPI-LAND + ASYEOS & & \\
n/p + n/ch & 48 $^{+12}_{-12}$ & 68$^{+13}_{-13}$\\
n/p + n/ch (p$_T$) & 74$^{+24}_{-24}$ & 91$^{+25}_{-25}$\\
\noalign{\smallskip}\hline
\end{tabular}
\end{center}
%\vspace*{2cm}  % with the correct table height
\end{table}

\begin{table}
\caption{Model dependence of the slope parameter $L$ due to uncertainties in
the values of the compressibility modulus $K_0$, in-medium effects on elastic
$NN$ cross-sections, Pauli blocking algorithm, value of the isovector neutron-proton
mass difference and scenario used for total energy conservation of the system.
The standard choice for parameters corresponds to entry labeled $K_0$=245 MeV. 
For other cases only the indicated parameter or scenario has been modified as mentioned.}
\tablab{lsym_mdi_moddep}       % Give a unique label
\begin{center}
\begin{tabular}{c|c|c|c|c}
\hline\noalign{\smallskip}
& \multicolumn{4}{|c}{L [MeV]}\\
\hline\noalign{\smallskip}
& \multicolumn{2}{|c|}{FOPI-LAND} & \multicolumn{2}{c}{ASYEOS} \\
\noalign{\smallskip}\hline\noalign{\smallskip}
Modified Parameter & n/p & n/H & n/ch & n/ch (p$_T$)\\
\noalign{\smallskip}\hline\noalign{\smallskip}
$K_0$=210 MeV & 85 & 103 & 57 & 80 \\
$K_0$=245 MeV & 80 & 94 & 39 & 65 \\
$K_0$=280 MeV & 73 & 81 & 34 & 60 \\
med cs & 81 & 97 & 48 & 60 \\
gPBA & 80 & 101 & 40 & 64 \\
GEC & 95 & 115 & 57 & 64 \\
$\delta^{*}_{n-p}(\rho_0,\beta$=0.5)=0.0 & 85 & 120 & 34 & 60 \\
$\delta^{*}_{n-p}(\rho_0,\beta$=0.5)=0.085 & 90 & 116 & 43 & 63\\
$\delta^{*}_{n-p}(\rho_0,\beta$=0.5)=0.28 & 83 & 95 & 39 & 72\\
\noalign{\smallskip}\hline
\end{tabular}
\end{center}
%\vspace*{2cm}  % with the correct table height
\end{table}

The values of npEFR and nhEFR measured by the FOPI-LAND collaboration have been compared in \figref{efrfopiland} with
theoretical predictions for different choices of certain model parameters or ingredients:
three values for the compressibility modulus $K_0$ (210, 245 and 280 MeV), vacuum versus
in-medium elastic $NN$ cross-sections and sPBA versus gPBA Pauli
blocking algorithms. The impact of these parameters on elliptic flow is sizable, see \figref{v2fopiland}.
By constructing elliptic flow ratios the model dependence is reduced substantially, particularly for npEFR (\figref{efrfopiland}).
Its impact on the extracted values of $L$ is presented in~\tabref{lsym_mdi_moddep} for both FOPI-LAND
and ASYEOS observables. Additionally, the impact of enforcing the conservation of total energy
(GEC scenario) is shown to lead to a stiffer value of $L$ by 10-20 MeV, with the exception of $p_T$ dependent nchEFR.
In total the estimated residual model dependence on $L$ amounts to 
18, 29, 22 and 17 MeV when extracted from npEFR, nhEFR, nchEFR and $p_T$ dependent nchEFR respectively.

\begin{figure*}
\begin{center}
\resizebox{0.45\textwidth}{!}{
\includegraphics{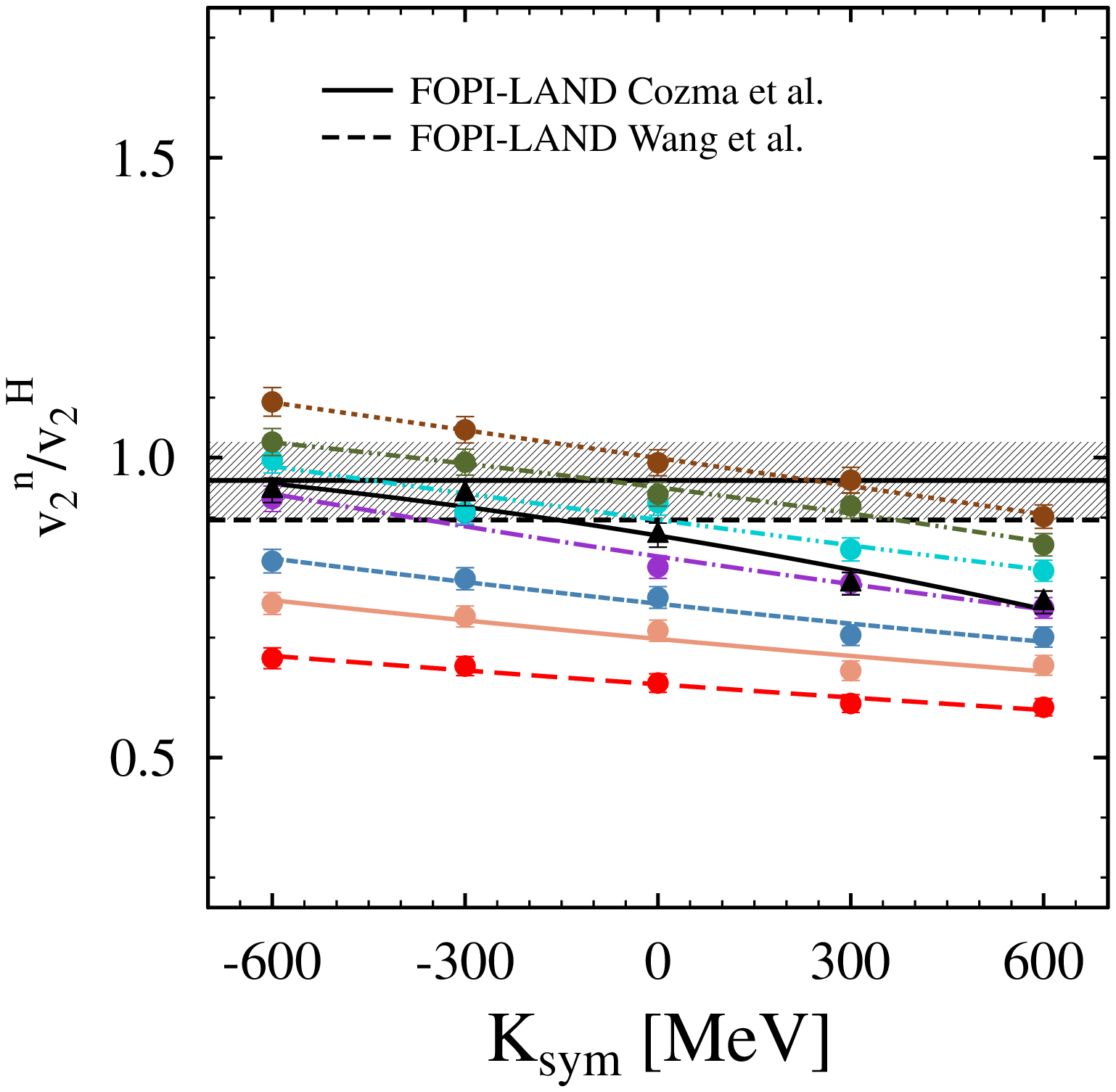}}
\resizebox{0.45\textwidth}{!}{
\includegraphics{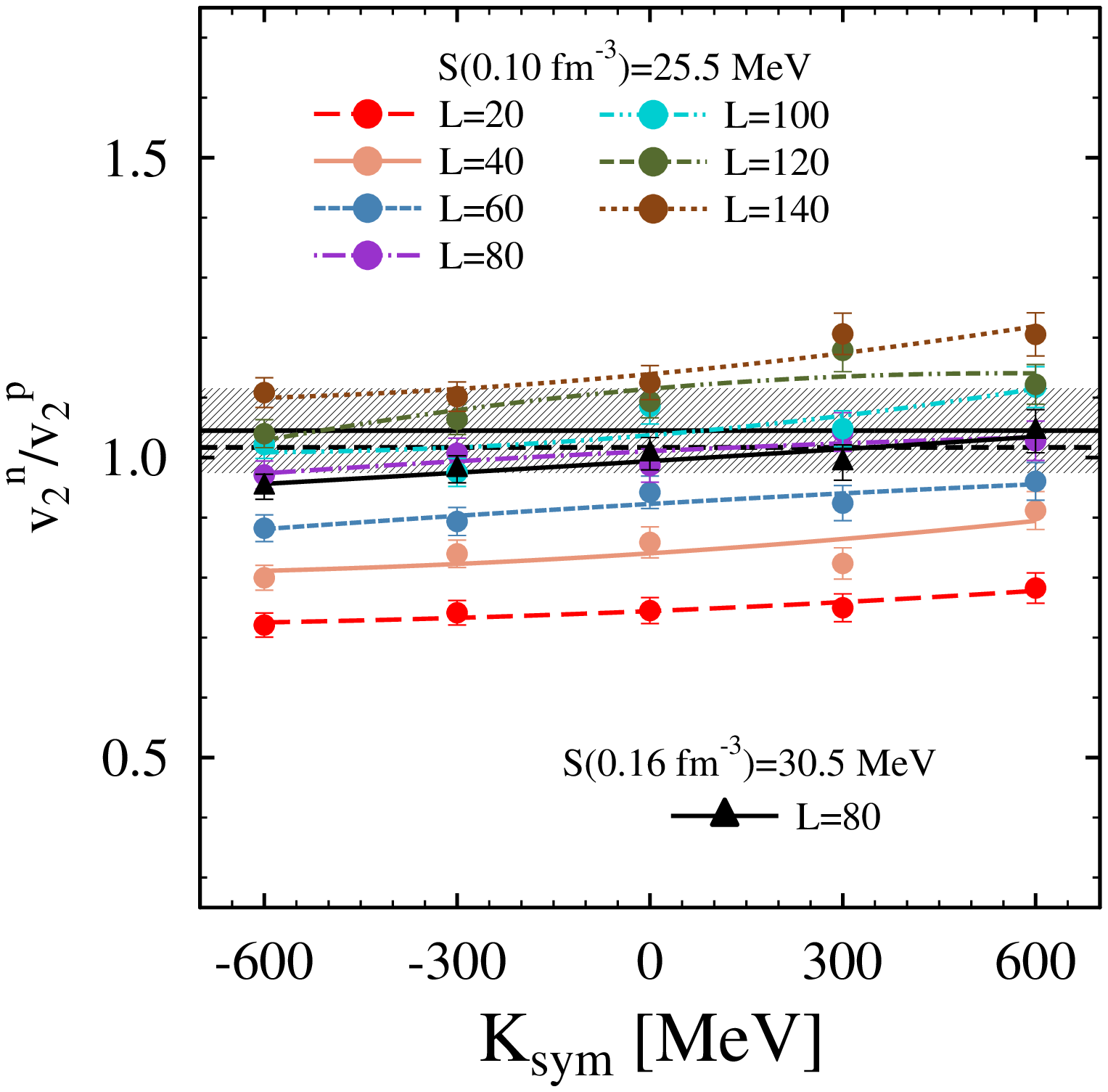}}
\end{center}
\caption{Neutron-to-hydrogen (left panel) and neutron-to-proton (right panel) elliptic flow ratios as
a function of curvature parameter $K_{sym}$ for selected values of the slope parameter $L$. 
Besides the standard choice S(0.10 fm$^{-3}$)=25.5 MeV, a calculation for a fixed value of the symmetry
energy at saturation, S(0.16 fm$^{-3}$)=30.5 MeV, is also shown. The same comments as in \figref{efrfopiland}
are in order for the experimental data.}
\figlab{efrksymsens}
\end{figure*}

In view of the results presented above it is concluded that the most reliable constraint
for the density dependence of the symmetry energy can currently be extracted by using
the npEFR observable. The following values for the slope and curvature of the SE are obtained
by taking the average of the two values listed in~\tabref{lsym_mdi_diffobs} and also taking
into account the residual model dependence summarized in~\tabref{lsym_mdi_moddep}:
\begin{eqnarray}
\eqlab{lkcmdi2_v1}
L&=&84\pm\phantom{1}30(\mathrm{exp})\pm18(\mathrm{th})\,\,\mathrm{MeV} \\
K_{sym}&=&30\pm142(\mathrm{exp})\pm85(\mathrm{th})\,\,\mathrm{MeV}\,. \nonumber
\end{eqnarray}
The experimental error includes the uncertainty due to the method used to analyze experimental data.
The theoretical error has been determined by adding in quadrature the deviations observed by modifying, within reasonable
limits, model parameters from their standard values (see~\tabref{lsym_mdi_moddep}. Systematical uncertainties due to the correlations
between $L$ and $K_{sym}$ have not been included and will be estimated in \secref{lksymmdi2}.
Additionally, possible contributions to the theoretical uncertainty due to the failure of the model
to accurately describe experimental multiplicities cannot be excluded. Their magnitude should be
however small given the good description of experimental flows of neutrons and protons. 

\subsection{MDI2 potential}
\seclab{lksymmdi2}
In this section we investigate the possibility of a simultaneous extraction of the values
of $L$ and $K_{sym}$ from a comparison of model predictions with available experimental data
for elliptic flow ratios. For that purpose the full MDI2 potential, that allows independent
adjustments of these two parameters, will be used in simulations of HIC.

Neutron-to-proton and neutron-to-hydrogen EFR have been shown to probe, on average, nuclear
matter with different values for density~\cite{Russotto:2016ucm}. Consequently,
the two observables are most sensitive to the density dependence of SE in regions of density
close to 1.5$\rho_0$ and $\rho_0$ respectively. It is expected that their sensitivity to
changing the value of $K_{sym}$, while keeping the slope parameter $L$ fixed, is different.

The results of such a calculation are shown in~\figref{efrksymsens} for nhEFR and npEFR.
Several simulations have been performed by keeping the value of $L$ fixed
while the curvature parameter has been varied in the 
interval -600 $ \le K_{sym} \le$ 600 MeV. The slopes of the nhEFR
and npEFR dependence on $K_{sym}$ are evidently different, the former being
negative while the latter is positive. The results are in agreement with the expectations
from the results on the average densities probed by these observables.
Consequently, the slope $L$ and curvature $K_{sym}$ parameters can be determined from
simultaneously comparing theoretical npEFR and nhEFR to experimental data.

\begin{figure*}
\begin{center}
\resizebox{0.45\textwidth}{!}{
\includegraphics{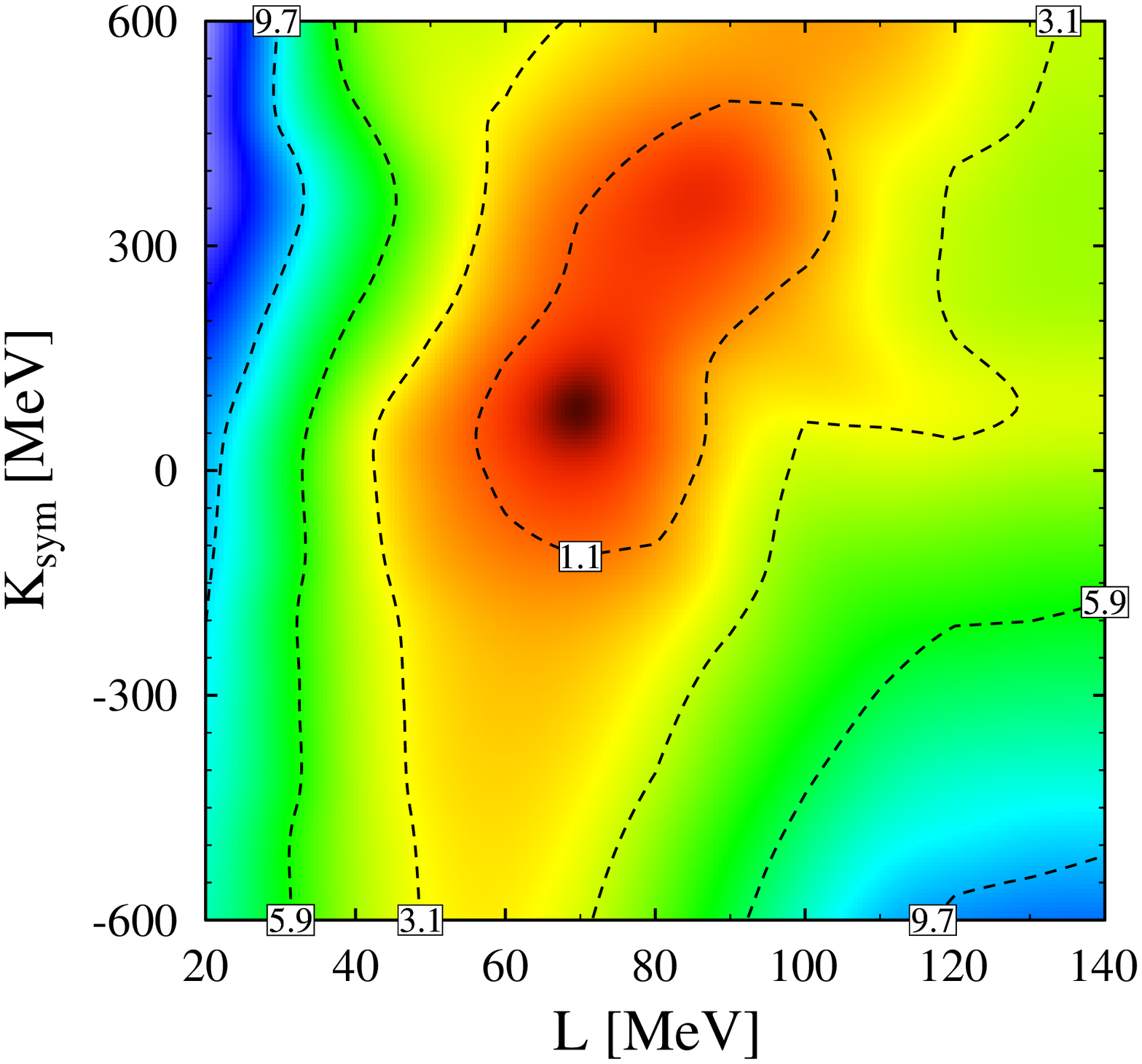}}
\resizebox{0.45\textwidth}{!}{
\includegraphics{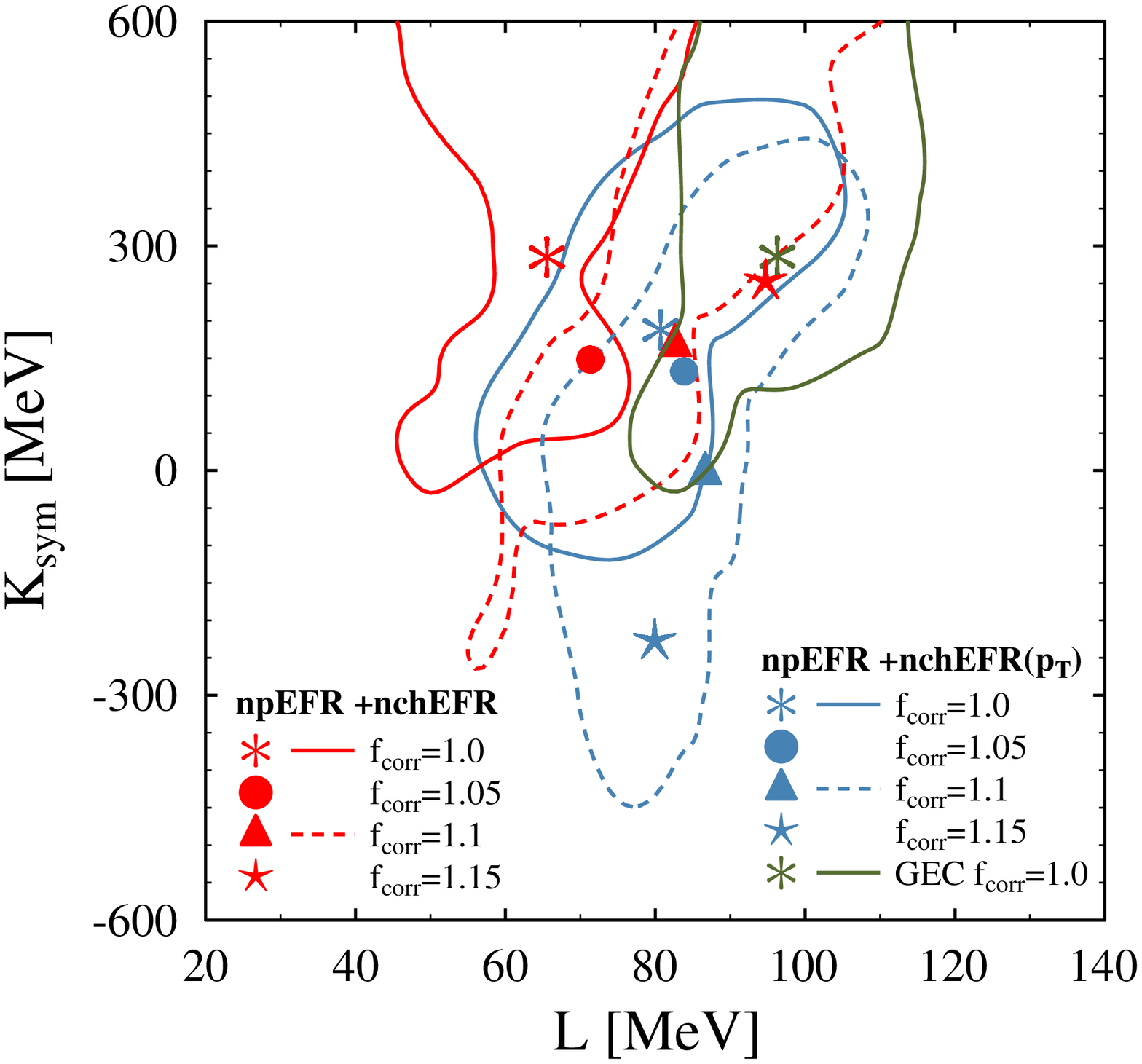}}
\end{center}
\caption{(Left panel) Constraint for the slope $L$ and curvature $K_{sym}$ extracted from
FOPI-LAND npEFR and ASYEOS $p_T$ dependent nchEFR. Curves for the 1, 2, 3 and 4 $\sigma$ confidence
levels are shown labeled by the corresponding $\chi^2/$point. For this case the comparison model-experiment
yields a minimum value for the goodness of fit parameter equal to $\chi^2/$point=0.14. (Right panel). Dependence of the
extracted values for the ($L$,$K_{sym}$) pair on the combination of observables used, (npEFR,nchEFR) versus
(npEFR, nchEFR($p_T$). For each case the impact of the parameter $f_{corr}$ used to correct for 
the systematical under-prediction of cluster-to-proton multiplicity ratios is shown. The constraint extracted
using the GEC scenario for total energy conservation is also presented. Contour curves correspond to 1 sigma
confidence levels.}
\figlab{fr_nchpt_systerr}
\end{figure*}

\begin{figure*}
\begin{center}
\resizebox{0.45\textwidth}{!}{
\includegraphics{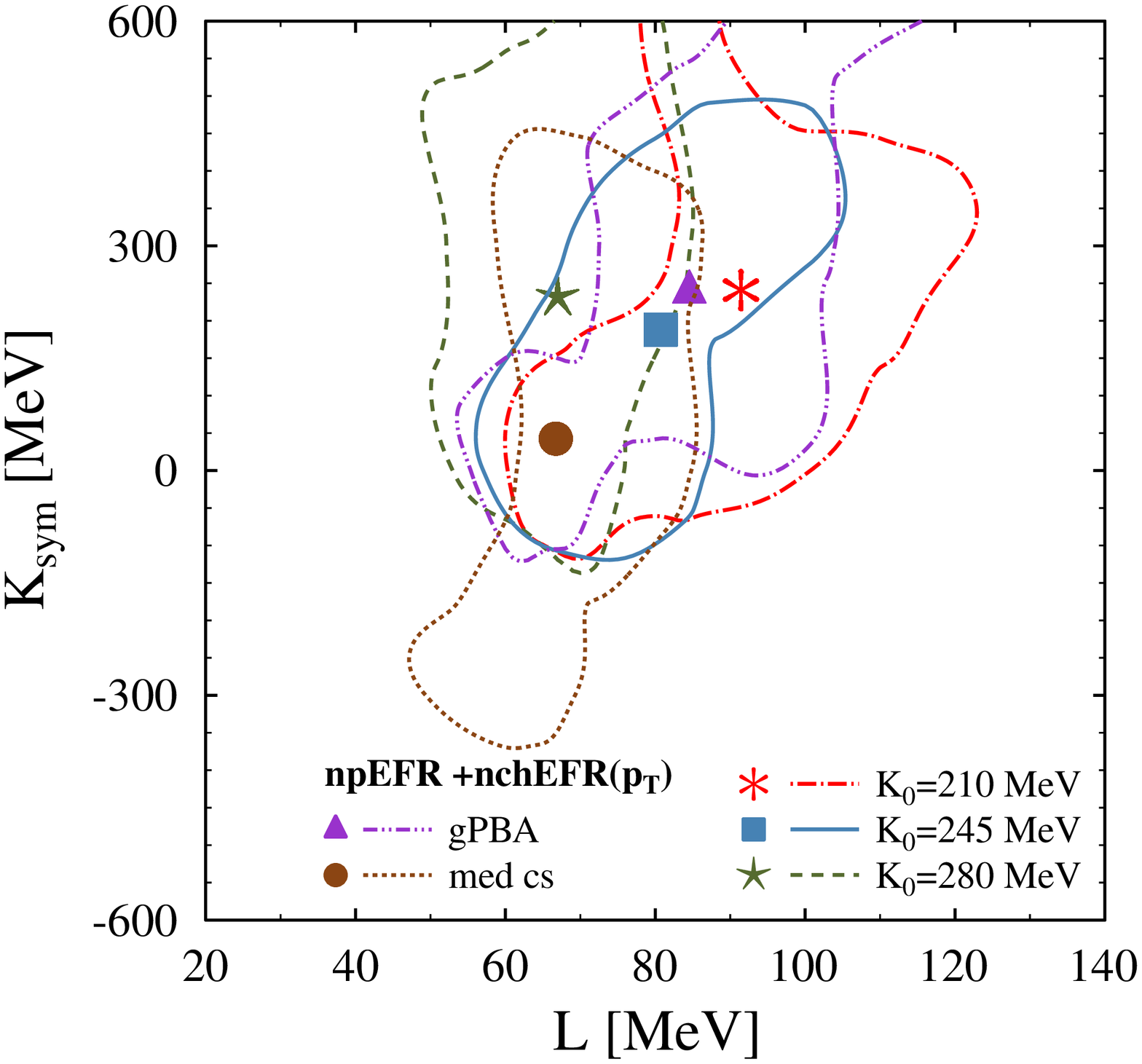}}
\resizebox{0.45\textwidth}{!}{
\includegraphics{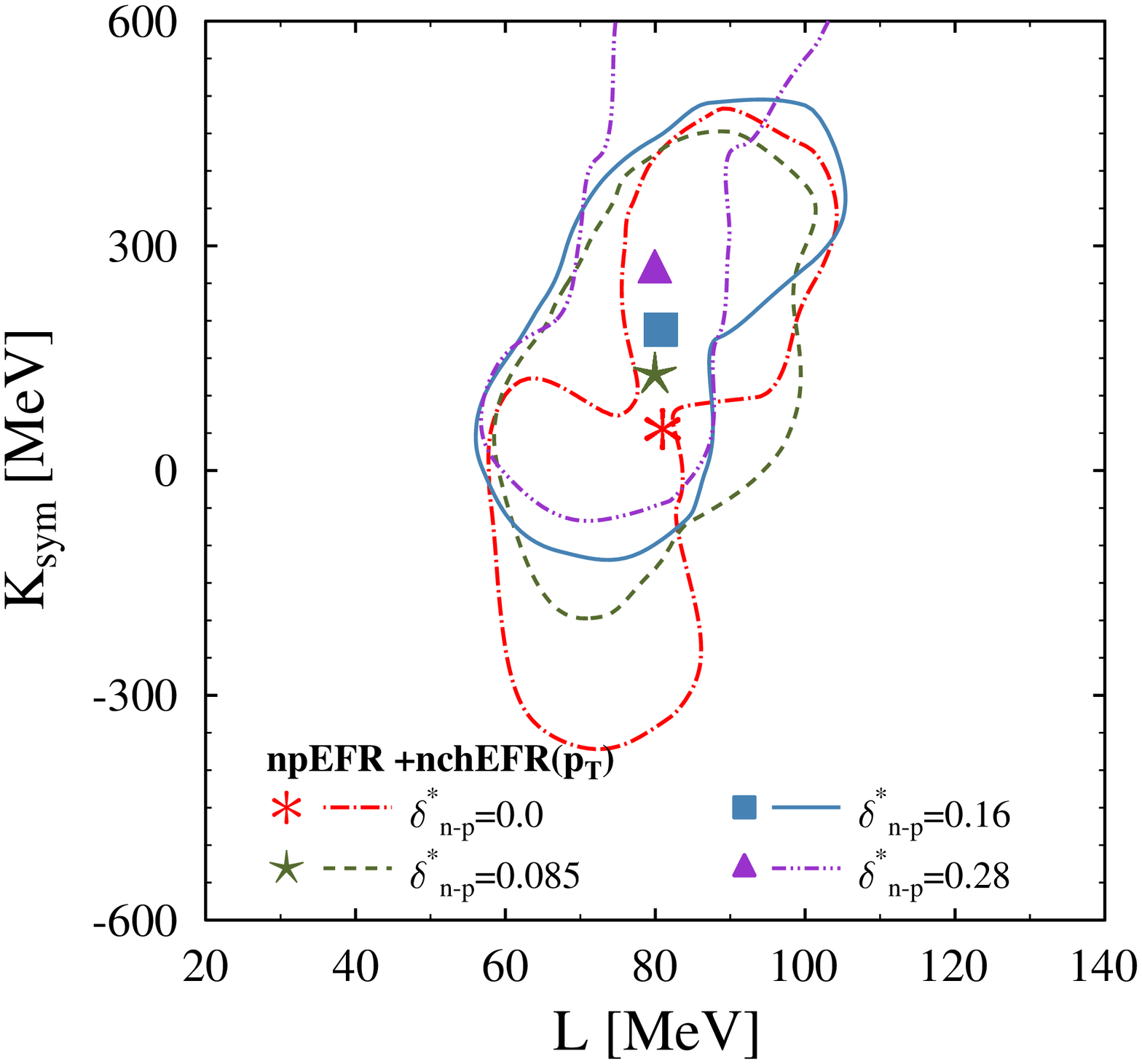}}
\end{center}
\caption{(Left panel) Dependence of the extracted values for $L$ and $K_{sym}$ on the compressibility
modulus $K_0$, in-medium effects on elastic $NN$ cross-sections and Pauli blocking algorithm. (Right Panel)
Impact of the value of the isovector nucleon-proton effective mass splitting on the extracted values for
$L$ and $K_{sym}$. The results correspond to the comparison theory-experiment using $\{$npEFR, nchEFR($p_T$)$\}$
set of observables. Only 1-$\sigma$ confidence limit contour curves are shown.}
\figlab{fr_nchpt_moddep}
\end{figure*}

In the case of the MDI2 potential the symmetry energy has been constrained to take a fixed value
at a sub-saturation point, $S(\rho$=0.10 fm$^{-3}$)=25.5 MeV. This leads to a dependence of the
value of the symmetry energy at saturation, $S_0$, on $K_{sym}$ for fixed values of $L$, as
was the case with the simulations presented in~\figref{efrksymsens}. For the case $L$=80 MeV
one has 31.8 $\le S_0 \le$ 39.6 MeV, the lower limit corresponding to $K_{sym}$=600 MeV. This
range of variation for $S_0$ includes the average favored value extracted from various terrestrial
and astrophysical measurements (see Refs.~\cite{Li:2013ola,Lattimer:2012xj} for a list of constraints
extracted from various sources) but it is several times wider than the reported
uncertainty. To test for a possible dependence
of EFR on the value of symmetry energy at saturation, a simulation with $S_0$=30.5 MeV 
and $L$=80 MeV has been performed. The result is shown in~\figref{efrksymsens} for both nhEFR
and npEFR (curve interpolating triangle points) and is found to be nearly identical to the
standard calculation for $L$=80 MeV. It is concluded that EFR in HIC of impact energies 
close to 400 MeV/nucleon are nearly insensitive to the value of the SE at saturation and consequently
this quantity cannot be constrained from such studies. 

An additional consequence of the noted sensitivity of npEFR to $K_{sym}$ is a systematic uncertainty
that affects the extracted value of $L$ from FOPI-LAND npEFR experimental data using the 
cMDI2 potential in the previous section. The cMDI2
potential differs from MDI2 by a potentially unrealistic constraint between $L$ and $K_{sym}$. Taking
this effect into account the constraint of~\eqref{lkcmdi2_v1} for the slope is corrected to 
\begin{eqnarray}
\eqlab{lkcmdi2_v2}
L&=& 84\pm30(\mathrm{exp})\pm 19(\mathrm{theor})\,\,\mathrm{MeV} 
\end{eqnarray}
The additional uncertainty has been added in quadrature to the theoretical component of the error and
has been estimated to amount to 6 MeV by assuming that the realistic value of the curvature parameter
lies in the range -300 $ \le K_{sym} \le$ 300 MeV. It is evident from~\figref{efrksymsens} that this
component of the uncertainty scales nearly linear with the assumed width of this range.

It turns out that the combination of experimental data for npEFR and nhEFR measured by the FOPI-LAND
collaboration is not precise enough to yield a constraint for $K_{sym}$ with a 1 $\sigma$ CL interval
within the -600 $ \le K_{sym} \le$ 600 MeV range. This can be achieved by replacing the FOPI-LAND nhEFR
data by either ASYEOS nchEFR or $p_T$ dependent nchEFR. The result for the latter choice
is shown in the left panel of \figref{fr_nchpt_systerr}. The extracted values for the slope and curvature
parameters read $L$=81$\pm$24 MeV and $K_{sym}$=188$\pm$307 MeV at 1 $\sigma$ CL. The 1, 2, 3 and 4 $\sigma$
CL contour curves are plotted, labeled by the corresponding value of $\chi^2/$point. The minimum value
for this quantity is reached for $L$=72 MeV and $K_{sym}$=78 MeV.

The favored values for the stiffness parameters for the combination of observables npEFR+nchEFR
can be read from the right panel of \figref{fr_nchpt_systerr}: $L$=66$\pm$20 MeV and $K_{sym}$=285$\pm$315 MeV.
(the upper limit for $K_{sym}$ is obviously underestimated). The values for the slope parameter $L$ extracted using
the two combination of observables are in very good agreement with each other. The curvature $K_{sym}$
extracted using npEFR + nchEFR is stiffer than the one obtained using the npEFR + nchEFR($p_T$) data set,
the two are however compatible at 1 $\sigma$ confidence level.
Results for npEFR+nchEFR by using the GEC scenario for total energy
conservation in the simulation of HIC are also presented. They are compatible at 1$\sigma$ level with the corresponding VEC case,
the extracted slope and curvature parameters being stiffer by $\Delta L$=17 MeV and $\Delta K_{sym}$=192 MeV respectively.

Additionally, the impact of correcting $v_2^{ch}$ due to systematic under-prediction of cluster-to-proton
multiplicity ratios, as prescribed by \eqref{v2corr}, on the extracted values for $L$ and $K_{sym}$
is also shown in the right panel of \figref{fr_nchpt_systerr}. Results for three values of the correction
factor are presented: $f_{corr}$=1.05, 1.10 and 1.15. Together with the uncorrected case, $f_{corr}$=1.0,
clear trends for the both npEFR + nchEFR and npEFR + nchEFR($p_T$) are evidenced. The magnitude and sign of
the changes of $L$ and $K_{sym}$ with $f_{corr}$ dependend strongly on the combinations of observable used.
Close values, well within the 1$\sigma$ level of uncertainty, for $L$ and  $K_{sym}$ are extracted for values
of the correction factor in the range 1.05 $\leq f_{corr} \leq$ 1.10. The central values of the two SE parameters
determined by averaging the results for the two combinations of observables and the most probable value
for the correction parameter, $f_{corr}$=1.10 (see \secref{thexpv2comp}), read: $L$=85 MeV and $K_{sym}$=95 MeV.

%Stiffer
%values for $L$ by $\Delta L \approx$ 15 MeV and softer ones for $K_{sym}$ by $\Delta K_{sym} \approx$ 350 MeV
%are determined for $f_{corr}$=1.15. Interestingly, for $f_{corr}$=1.10 the extracted constraints for the two combinations
%of observables are closest in parameter space ($L \approx$ 65 MeV, $K_{sym} \approx$ 0 MeV), 
%well within the 1 $\sigma$ CL.

In \figref{fr_nchpt_moddep} the model dependence of the extracted constraint for the SE from npEFR+nchEFR($p_T$)
combination of observables is studied. In the left panel the impact of variation of the compressibility
modulus in the range 210 $\le K_0 \le$ 280 MeV, vacuum versus in-medium elastic $NN$ cross-sections and 
sPBA versus gPBA Pauli blocking algorithms is presented.

A correlation between the extracted value of $L$ with the used value of $K_0$ is observed,
%although with a rather small statistical significance.
a stiffer isoscalar EoS leading to a softer slope of the SE. 
The value of $K_{sym}$ appears as uncorrelated to the compressibility modulus of symmetric nuclear matter
or the slope $L$. This contrast the correlation with positive slope between $L$ and $K_{sym}$ built in 
the original MDI potential~\cite{Das:2002fr} (and mimicked by cMDI2 potential) and other potentials based
on the Skyrme interaction~\cite{Chen:2011ib,Vidana:2009is,Ducoin:2011fy}. It provides a posteriori support
for the introduction of the extra parameter $y$ in the expression of the MDI2 potential and can
be used to constrain the lesser known terms of the short-range $NN$ interaction.
Nevertheless, this latter finding bears a rather small statistical significance.
%~\cite{Gandolfi:2011xu,Steiner:2011ft}.

The impact on the extracted values
of these two parameters by varying $K_0$ in the mentioned interval amounts to $\Delta L \approx$ -30 MeV and
$\Delta K_{sym} \approx$ 75 MeV. It induces uncertainties in the extrapolated values of the SE of 8, 19 
and 35 MeV at 2, 3 and 4$\rho_0$ respectively. The trends for $L$ are the same as in the case of the cMDI2 potential, see
\tabref{lsym_mdi_moddep}, but larger in magnitude. A constraint of the
SE extracted from heavy-ion data that allows precise extrapolations at densities encountered at the center
of neutron stars will therefore require also precise quantitative determination of the density dependence of the
isoscalar part of the EoS, namely the compressibility $K_0$ and the skewness parameter $J_0$. 

In-medium effects on the elastic $NN$ cross-sections favor, on average, a moderately softer values for $L$ and $K_{sym}$. 
Choosing the gPBA Pauli blocking algorithm leads to a slightly stiffer $L$ and $K_{sym}$ as compared 
to the standard sPBA option.

The impact of changing the value of the isovector neutron-proton mass difference $\delta m^*_{n-p}$ is presented
in the right panel of \figref{fr_nchpt_moddep}. A correlation between $\delta m^*_{n-p}$ and the stiffness of SE is noticed
only for the $K_{sym}$, a higher value of former quantity favors a stiffer $K_{sym}$ albeit with a small statistical significance.

The results presented in this section are summarized by the following constraints for the slope $L$ and curvature
$K_{sym}$ parameters
\begin{eqnarray}
\eqlab{lkmdi2}
L&=&85\pm\phantom{3}22(\mathrm{exp})\pm\phantom{2}20(\mathrm{th})\pm\phantom{1}12(\mathrm{sys})\,\,\mathrm{MeV} \\
K_{sym}&=&96\pm315(\mathrm{exp})\pm170(\mathrm{th})\pm166(\mathrm{sys})\,\,\mathrm{MeV}\,. \nonumber 
\end{eqnarray}
The indicated uncertainties are of experimental, theoretical (model dependence) and systematical (underprediction
of cluster-to-proton multiplicity ratios) origin. The theoretical uncertainty has been determined by adding in quadrature
all model dependence illustrated in \figref{fr_nchpt_systerr} and \figref{fr_nchpt_moddep}. The quoted value for the
systematical error for $K_{sym}$ has been estimated as the half difference between the maximum
and minimum average values for $K_{sym}$ when the $f_{corr}$ parameter is varied in the range [1.05,1.15], since 
the case $f_{corr}$=1.0 is clearly unrealistic. The value of $L$ is in good agreement with the corrected value
$L_{corr}$ extracted using the cMDI2 potential and the same combination of observables (see bottom lines of
\tabref{lsym_mdi_diffobs}).

\begin{figure*}
\begin{center}
\resizebox{0.45\textwidth}{!}{
\includegraphics{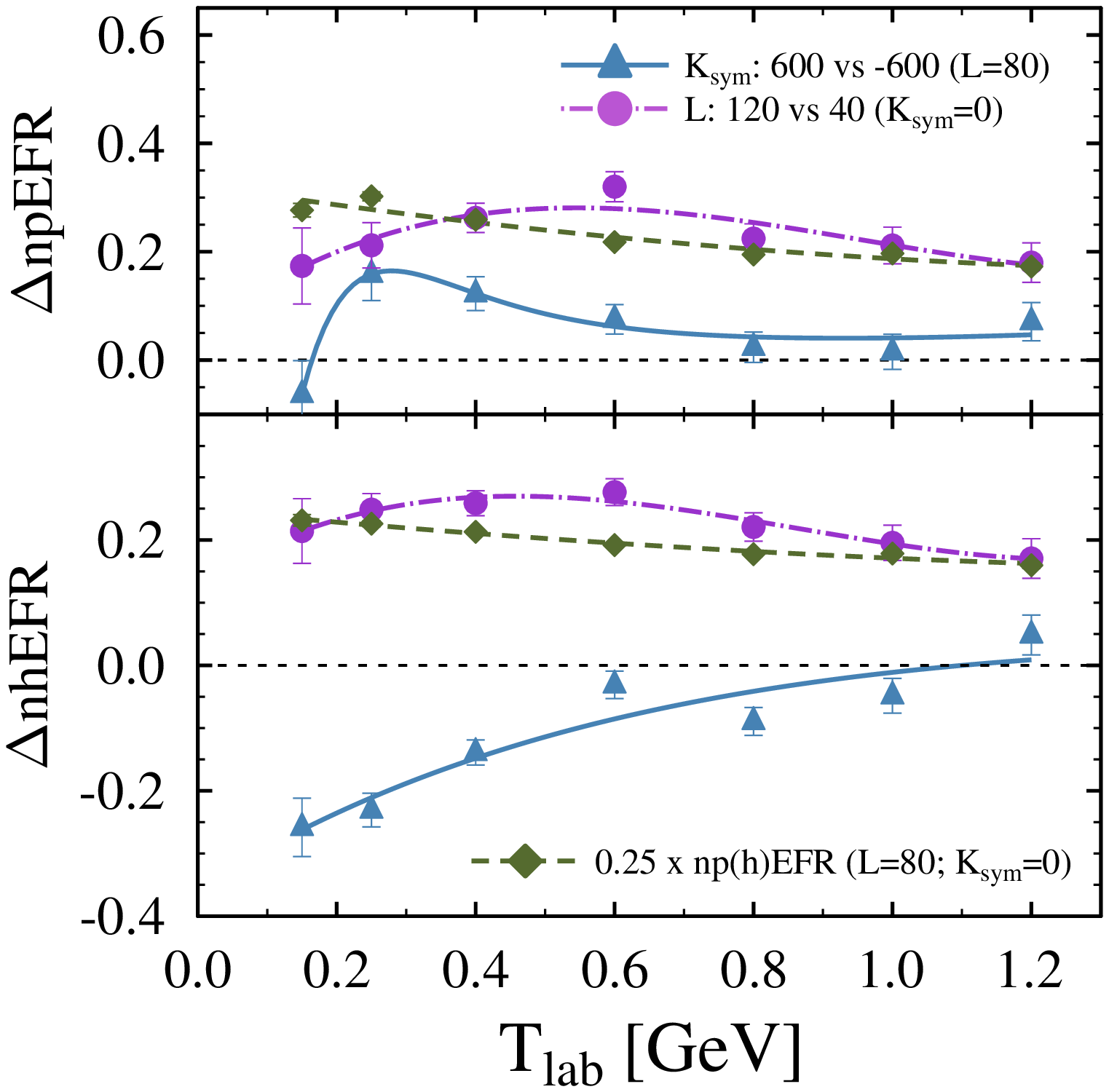}}
\resizebox{0.45\textwidth}{!}{
\includegraphics{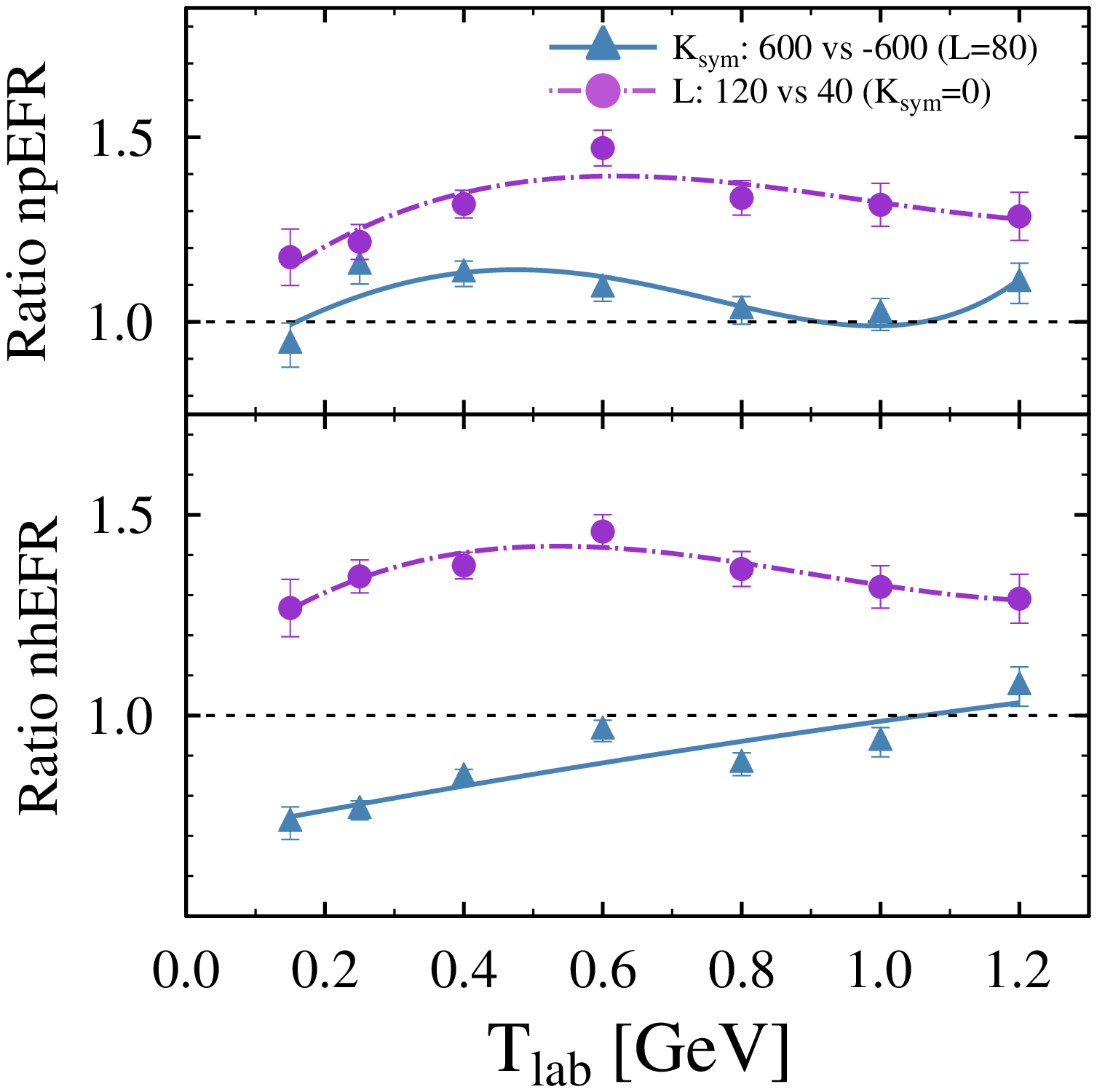}}
\end{center}
\caption{(Left panel) Sensitivity of the npEFR (top) and nhEFR (bottom) to the slope $L$ and 
curvature $K_{sym}$ as a function of the projectile kinetic energy for $^{197}$Au+$^{197}$Au collisions.
The plotted quantity is defined as the difference of predictions of the chosen observable (npEFR or nhEFR)
between a stiff and a soft asy-EoS. For the slope $L$ it is defined as
np(h)EFR($L$=120;$K_{sym}$=0)-np(h)EFR($L$=40;$K_{sym}$=0). Similarly for the curvature
$K_{sym}$ the definition np(h)EFR($L$=80;$K_{sym}$=600)-np(h)EFR($L$=80;$K_{sym}$=-600) has been used.
The energy dependence of np(h)EFR($L$=80;$K_{sym}$=0), scaled by a factor of 0.25, is also shown. Quoted values
of $L$ and $K_{sym}$ are in units of MeV.} (Right panel) The sensitivity of npEFR (top) and nhEFR (bottom)
to the slope $L$ and $K_{sym}$ as a function of projectile kinetic energy for $^{197}$Au+$^{197}$Au collisions
defined as ratios of the mentioned observables for a stiff and a soft asy-EoS. The same combintations of $L$ and $K_{sym}$
as for the left panel have been used to determine the plotted ratios.
\figlab{efrenerdep}
\end{figure*}

\section{Discussion and Outlook}
\seclab{discussion}
The slope $L$ and curvature $K_{sym}$ of the SE enter in the expression of the isospin dependent component of nuclear matter
compressibility,
\begin{eqnarray}
 K_\tau=K_{sym}-6L-\frac{J_0}{K_0}L\,,
\end{eqnarray}
where corrections of order $\delta^2$ or higher have been neglected~\cite{Chen:2009wv}. This quantity
is usually extracted from experimental data of neutron skin sizes~\cite{Centelles:2008vu}, location of
the centroid of isoscalar giant resonances~\cite{Shlomo:1993zz,Li:2010kfa,Stone:2014wza}, isospin diffusion
in HIC~\cite{Chen:2004si} or determined theoretically using as input effective potentials of the Skyrme or Gogny type
constrained by nuclear data~\cite{Yoshida:2006nk,Sagawa:2007sp,Chen:2009wv}. The determined values for $K_\tau$
generally agree with each other but the claimed accuracies vary significantly: 
$K_\tau$=-500$\pm$50 MeV~\cite{Sagawa:2007sp}, $K_\tau$= -500$\pm$100 MeV~\cite{Li:2010kfa}, 
$K_\tau$=-370$\pm$120 MeV~\cite{Chen:2009wv}, $K_\tau$=-500$^{+150}_{-100}$ MeV~\cite{Centelles:2008vu}
and $K_\tau$=-600$\pm$250 MeV~\cite{Stone:2014wza}. In Ref.~\cite{Shlomo:1993zz} the authors determine directly
$K_{sym}$, the obtained values and accuracies depend strongly on the data set and fitting procedure.

The constraints derived in \secref{lksymcmdi2} and \secref{lksymmdi2} for the cMDI2 and MDI2 potentials
lead to the following values for $K_\tau$ respectively
\begin{eqnarray}
 K_\tau=-354 \pm 228 \mathrm{\phantom{a}MeV} (\mathrm{cMDI2}) \\
 K_\tau=-290 \pm 421 \mathrm{\phantom{a}MeV} (\mathrm{MDI2})\,.
\end{eqnarray}
The result extracted using the cMDI2 potential is almost the same as that of Ref.~\cite{Chen:2009wv}. This
is due to a very similar and potentially unphysical constraint between $L$ and $K_{sym}$
enforced by both the cMDI2 and MDI Gogny potentials. Their energy dependence is however different
at high values of momentum. This, together with the different values for $K_0$, 
induces modifications of the extracted value for $L$ of about 10 MeV,
as can be seen by comparing the results in \secref{prevmodel} and \secref{lksymcmdi2}, leaving $K_\tau$ almost unaffected.
The uncertainty of cMDI2 extracted $K_\tau$ is at the upper limit of those quoted in the literature.

The density dependence of SE favored by the full MDI2 is somewhat stiffer and consequently the value of $K_\tau$ is closer
to zero, the central value lying outside of the 3$\sigma$ CL interval favored by the most accurate extractions in the
literature quoted above. The MDI2 constraint is two times less accurate than the cMDI2 one. Clearly, this
is entirely due to lifting the constraint between $L$ and $K_{sym}$. This conclusion casts some doubt at least
on the validity of the reported accuracies for some of the extracted values of $K_\tau$ in the literature. 
More careful analyses need to be performed on whether the so far employed observables are 
in fact sensitive to $K_\tau$ (not just $L$) and that the reported results are not biased by the parametrization
used for the symmetry energy.

Up to 1.5$\rho_0$ the density dependence of the SE is dominated by the slope term. Consequently
the constraints extracted using cMDI2 and MDI2 potentials are of comparable accuracies in this region.
However, the allowed range for $K_{sym}$ extracted in this work is clearly not precise enough for the purpose
of extrapolating the SE at densities above 1.5$\rho_0$. The largest contribution to the determined
uncertainty originates from experimental data, particularly the FOPI-LAND npEFR. An improvement of the
experimental relative accuracy for this observable, comparable to that achieved by the ASYEOS
collaboration for nchEFR, would lead to a decrease of the experimental uncertainty of $K_{sym}$
to an estimated value of 200 MeV. Further improvements may be possible if heavy-ion collisions are studied
experimentally at bombarding energies different from 400 MeV/nucleon.

To investigate for this possibility, the sensitivity of npEFR and nhEFR to $L$ and $K_{sym}$ has been determined
for $^{197}$Au + $^{197}$Au collisions at projectile energies between 150 and 1200 MeV/nucleon and
impact parameter $b \le$ 7.5 fm. The results, restricted to the FOPI-LAND geometry and filter, are presented
in~\figref{efrenerdep}. The sensitivity of np(h)EFR to $L$ reaches a maximum around 600 MeV/nucleon and
decreases towards both higher and lower energies. In the case of $K_{sym}$ the maximum of sensitivity of the npEFR
observable is reached close to 250 MeV/nucleon impact energy and then it decreases as $v_2$ changes sign at
lower impact energies~\cite{Andronic:2006ra}. For nhEFR the sensitivity is seen to increase monotonically towards
lower impact energies. This is due to the fact of probing, on average, ever lower subsaturation densities. In this
context the correlations of possible non-negligible higher order terms in the Taylor expansion of
the SE (see \eqref{taylorexpse}) with $L$ and $K_{sym}$ induced by the parametrization chosen for
the EoS may lead to a bias in the extracted values for these two parameters. As experimental data will become more
accurate this issue will need to be investigated. A similar observation is in order if densities
far above satuation are probed as is the case of the npEFR observable at the upper end of the interval of impact energies.

It has been investigated how this conclusion is affected if, instead of determininng the coalescence
model parameters from fits to experimental multiplicities, independent values of impact energy and asy-EoS stiffness are chosen
for them. To that end the relation $\delta r_{pp}$=$\delta r_{np}$=$\delta r_{nn}$=$\delta r$ has been enforced.
Calculations with values in the ranges $\delta r$=2.0-4.0 fm and $\delta p$=0.2-0.3 GeV/c have been
performed. A non-negligible dependence of the sensitivity of EFR to $L$ and $K_{sym}$ on the values of
the coalescence parameters has been observed in the low energy region. 
%In particular the magnitude of the sensitivy the peak close to $T_{lab}$=250 MeV/nucleon for the npEFR $K_{sym}$
%sensitivity is visibly affected.
However, the conclusions stated above remain qualitatively the same.

It thus becomes apparent that experimental measurements of EFR at projectile energies
in the neighborhood of 250 MeV/nucleon are best suited for a more precise determination of $K_{sym}$. Based on the calculated
sensitivities in~\figref{efrenerdep} and those described in the paragraph above it is estimated that $K_{sym}$ could
be extracted with an accuracy in the neighborhood of 150 MeV.

This result challenges the expectation that using higher projectile energies is better suited for the extraction
of the density dependence of symmetry energy using flow observables. This view is based on the 
well known fact that the maximum density probed in heavy-ion collisions increases
with impact energy, namely from 1.75$\rho_0$ at 200 MeV/nucleon to 3.25$\rho_0$ at 2.0 GeV/nucleon for
the case of Sn+Sn mid-central collisions \cite{Li:2002yda}.
The lifetime of the high-density fireball does however decrease quite rapidly towards higher impact energies.
The sensitivity of an observable to the high density EoS is therefore the result
of these two competing features that have opposite effects.

\section{Summary}
A QMD transport model that employs a modified momentum dependent interaction (MDI2) potential, supplemented by
a phase-space coalescence model fitted to experimental multiplicities has been used to study the
density dependence of the symmetry energy above the saturation point by a comparison with 
experimental neutron-to-proton, neutron-to-hydrogen and neutron-to-charged particles elliptic
flow ratios (EFR) in $^{197}$Au+$^{197}$Au heavy-ion collisions at 400 MeV/nucleon impact energy.
It has been recognized that for a trustworthy extrapolation of symmetry energy stiffness
constraints to high density, both the slope $L$ and curvature $K_{sym}$ parameters need to be
determined accurately. Such a study has been suggested to be possible by a calculation, using the 
same transport model, that has revealed that neutron-to-proton and neutron-to-hydrogen (or alternatively
neutron-to-charged particles) EFR probe on average different densities~\cite{Russotto:2016ucm} at this impact energy.

To that end, a Gogny interaction inspired MDI potential~\cite{Das:2002fr} has been modified, by the addition
of an extra density dependent - momentum independent term, to allow for independent variations of the slope $L$
and curvature $K_{sym}$ parameters. The particular choice for this extra term also allows independent
modifications of the isovector neutron-proton effective mass difference, whose value is currently
a topic of hot debate. The momentum dependent part of the potential has been adjusted to reproduce the
empirical nucleon optical potential~\cite{Arnold:1982rf,Hama:1990vr,Cooper:1993nx} following the procedure outlined in
Ref.~\cite{Xu:2014cwa}.

The final state spectra of heavy-ion collisions have been determined by making use of a
minimum spanning tree coalescence algorithm that recognizes all clusters with $A \le$ 15 and a few
additional ones of higher mass and $Z \le$ 8. All clusters with lifetimes larger than 1 ms are considered as stable,
the rest are decayed via strong-interaction channels until a stable daughter is reached. The $r$-space coalescence
parameters are assumed to be isospin dependent and are determined from fits to experimental light cluster
multiplicities for each impact energy of interest, up to $Z \le$ 6 where available~\cite{Reisdorf:2010aa}. 
It is observed that multiplicities of free neutrons and protons are systematically over-predicted, 
those for $^3$H, $^3$He, $^4$He , Li are under-predicted while $^2$H, Be, B, C nuclei are generally in good agreement with experiment.

Theoretical transverse and elliptic flows of protons and light clusters (taken separately) generally reproduce well 
the corresponding FOPI experimental data~\cite{FOPI:2011aa}. It is however noticed that theoretical elliptic flows
of charged particles systematically under-predict the ASYEOS experimental data. This can be traced back to
the underprediction of experimental light-cluster-to-proton multiplicity ratios, a systematic effect
that also impacts elliptic flow of hydrogen values. Multiplicative correction factors that amount to approximately
1.1 and 1.075 respectively, with a 5$\%$ uncertainty, have been estimated for these two cases. In principle
such effects can be avoided by developing or using existing transport models that include light-cluster
degrees of freedom~\cite{Danielewicz:1991dh} or compare theoretical and experimental
coalescence invariant results~\cite{Famiano:2006rb}.

The different sensitivity of neutron-to-proton and neu-tron-to-hydrogen EFR to the value of $K_{sym}$ has been
proven by fixing the value of the slope $L$ and varying the curvature in the interval 
-600 MeV$\le K_{sym} \le$ 600 MeV. Consequently the following constraint for the density dependence of
the symmetry energy has been extracted from a comparison with experimental FOPI-LAND neutron-to-proton and ASYEOS
neutron-to-charged particles EFR data
\begin{eqnarray}
L&=&85\pm\phantom{3}22(\mathrm{exp})\pm\phantom{2}20(\mathrm{th})\pm\phantom{1}12(\mathrm{sys})\,\,\mathrm{MeV} \nonumber\\
K_{sym}&=&96\pm315(\mathrm{exp})\pm170(\mathrm{th})\pm166(\mathrm{sys})\,\,\mathrm{MeV}\,. \nonumber 
\end{eqnarray}
Theoretical errors include effects due to uncertainties in the isoscalar part of the equation of state,
value of the isovector neutron-proton effective mass splitting, in medium effects on the elastic nucleon-nucleon
cross-sections, Pauli blocking algorithm variants and scenario considered for the conservation of the total energy of the system.
Systematical uncertainties are generated by the inability of the transport model to reproduce light-cluster-to-proton
multiplicity ratios. The extracted value for $L$ is in agreement with constraints extracted from other studies
and of comparable total uncertainty. The value of $K_{sym}$ is however imprecise and the extracted symmetry energy
cannot be extrapolated accurately above 1.5 saturation density. 

It has been shown that a much more precise constraint for $K_{sym}$ and the isospin dependent component
of the nuclear matter compressibility $K_\tau$ can falsely be reported if potentials that include
potentially unphysical correlations between their parameters are employed. Such existing constraints for the
latter quantity should be revisited by studying in more depth the model dependence induced by these type
of correlations. The correlation between $L$ and $K_{sym}$ extracted from models based on the Skyrme or
Gogny interactions~\cite{Vidana:2009is,Ducoin:2011fy,Chen:2011ib} is not favoured by the findings of the
present study. The difference is however of a rather small statistical significance.

A constraint for $L$, free of the mentioned systematical uncertainties, can be extracted from the
FOPI-LAND neutron-to-proton EFR alone. After correcting for the sensitivity of this observable to $K_{sym}$ and
considering the same possible model dependence sources as before, the constraint
\begin{eqnarray}
L&=& 84\pm30(\mathrm{exp})\pm 19(\mathrm{theor})\,\,\mathrm{MeV} \nonumber
\end{eqnarray}
is extracted. The experimental error includes also uncertainties due to different possible approaches to
analyze the experimental data. The difference with respect to constraints for $L$ extracted
using a previous version of the model~\cite{Cozma:2013sja} and the same experimental data have been shown to be the result of an 
unsatisfactory description of nuclear density profiles, the use of a density cut-off method to determine
free nucleon spectra in that study and the isospin dependence of the Pauli blocking algorithm.

Finally, an analysis of the dependence of the sensitivity of EFR to $K_{sym}$ on the energy
of the projectile nucleus has been performed. It has been shown that, contrary to expectations, a lower
impact energy of about 250 MeV/nucleon for $^{197}$Au+$^{197}$Au heavy-ion collisions is most suitable
for constraining the curvature parameter $K_{sym}$. This is the result of two competing effects with
opposite impact: higher impact energies lead to higher probed central densities but with a lower
lifetime of the fireball~\cite{Li:2002yda}. The maximum sensitivity for $L$ is reached around 600 MeV/nucleon impact energies,
but in this case the energy dependence is weaker than for $K_{sym}$.

\section{Acknowledgments}
The research of M.D.C. has been financially supported by the Romanian Ministry of Research through 
Contract No. PN 16420101/2017. The assistance of the DFCTI department of IFIN-HH with 
maintenance of the computing cluster on which part of the simulations were performed 
is gratefully acknowledged. The author is indebted to the referees for valuable comments.

%\begin{figure}
%\resizebox{0.75\textwidth}{!}{%
%  \includegraphics{leer.eps}
%\vspace{5cm}       % Give the correct figure height in cm
%\end{figure}
%\begin{figure*}
%\vspace*{5cm}       % Give the correct figure height in cm
%\end{figure*}
%

\section*{Appendix A}
\seclab{appendix}
The analytical expressions of the integrals over one and two Fermi spheres that
appear in the formulae for the single-particle potential and equation of states respectively for the case of
cold nuclear matter are provided in this Appendix. They are needed for fixing the parameters appearing
in \eqref{eos}. The derivations can be performed using only elementary methods, but are rather tedious.
The result for the integral appearing in \eqref{sympot} reads~\cite{Das:2002fr}
\begin{eqnarray}
 &&I_1(p_F(\tau))=\int d^{\!\:3} \vec{p}\!\;'\, \frac{f_\tau(\vec{r},\vec{p}\!\;')}{1+(\vec{p}-\vec{p}\!\;')^2/\Lambda^2} \\
&&=\frac{2\pi}{h^3}\Lambda^3\Bigg [ \frac{\Lambda^2+p_F^2(\tau)-p^2}{2\Lambda p}\,\mathrm{ln}\,\frac{\Lambda^2+[p+p_F(\tau)]^2}
{\Lambda^2+[p-p_F(\tau)]^2} \nonumber\\
&&+\frac{2p_F(\tau)}{\Lambda}+ 2\,\Bigg ( \mathrm{arctan} \frac {p-p_F(\tau)}{\Lambda} 
-\mathrm{arctan} \frac {p+p_F(\tau)}{\Lambda} \Bigg ) \Bigg ]. \nonumber
\end{eqnarray}
The analytical expression for the two-Fermi-spheres integral appearing in \eqref{eos} is\footnote{The expression for the same integral
presented in Ref.~\cite{Das:2002fr} is inexact. }
\begin{eqnarray}
\eqlab{int2fermispheres}
 &&I_2(p_F(\tau),p_F(\tau'))=\int\!\!\!\!\int d^{\!\:3} \vec{p}\,d^{\!\:3} \vec{p}\!\;'
\frac{f_\tau(\vec{r},\vec{p}) f_{\tau'}(\vec{r},\vec{p}\!\;')}{1+(\vec{p}-\vec{p}\!\;')^2/\Lambda^2} \\
&&=16\pi^2 \frac{\Lambda^3}{h^6}\Bigg [  \frac{1}{2}\frac{p_F(\tau)p_F(\tau ')}{\Lambda} 
\Big [\,p_F^2(\tau)+p_F^2(\tau ')-\frac{1}{3}\Lambda^2 \Big]  \nonumber\\
&&\phantom{a}+\frac{2}{3} \big [\,p_F^3(\tau)-p_F^3(\tau ') \big ]\,\mathrm {arctan} \,\frac{p_F(\tau)-p_F(\tau ')}{\Lambda} \nonumber\\
&&\phantom{a}-\frac{2}{3} \big [\,p_F^3(\tau)+p_F^3(\tau ') \big ]\,\mathrm {arctan}\,\frac{p_F(\tau)+p_F(\tau ')}{\Lambda} \nonumber\\
&&\phantom{a}+\Big\{-\frac{1}{8} \frac{[\,p_F^2(\tau)-p_F^2(\tau ')]^2}{\Lambda}+\frac{1}{4}\Lambda\,[\,p_F^2(\tau)-p_F^2(\tau ')] \nonumber\\
&&\phantom{a}+\frac{1}{24}\Lambda^3 \Big\}\,\mathrm{ln}\,\frac{\Lambda^2+[\,p_F(\tau)+p_F(\tau ')]^2}
{\Lambda^2+[\,p_F(\tau)-p_F(\tau ')]^2}
\Bigg ] \nonumber
\end{eqnarray}
In the process of fixing the parameters of the potential the contribution of the previous integral to both 
the EoS of symmetric nuclear matter and symmetry energy, \eqref{eos}, will also be needed. The former is 
easily found by setting $p_F(\tau)=p_F$ and $p_F(\tau ')=p_F$ in the above expression, 
with $p_F=(3\pi\rho/2)^{1/3}$ being the Fermi momentum of symmetric nuclear matter of density $\rho$.
The latter is found by expanding \eqref{int2fermispheres} in powers of the isospin asymmetry $\beta$. 
After a tedious calculation, the coefficient of $\beta^2$ is determined to be
\begin{eqnarray}
 &&\int\!\!\!\!\int d^{\!\:3} \vec{p}\,d^{\!\:3} \vec{p}\!\;'
\frac{f_\tau(\vec{r},\vec{p}) f_{\tau'}(\vec{r},\vec{p}\!\;')}{1+(\vec{p}-\vec{p}\!\;')^2/\Lambda^2}=16\pi^2 \frac{\Lambda^3}{h^6}\,
\sum_{n=0}^{\infty} S^{(n)}_{\tau ,\tau'}\,\beta^n \nonumber\\
&&S^{(2)}_{\tau,\tau'}=\frac{3-\tau\tau'}{9}\frac{p_F^4}{\Lambda}+\frac{2-\tau\tau'}{54}\Lambda p_F^2+
\frac{8}{27}\frac{\Lambda p_F^4}{\Lambda^2+4p_F^2} \nonumber\\
&&\phantom{aaaaaa}+\frac{16}{27}(1+\tau\tau')\frac{\Lambda p_F^6}{(\Lambda^2+4p_F^2)^2} \nonumber\\
&&\phantom{aaaaaa}-\frac{1}{18}\frac{p_F^2}{\Lambda}\big[ \Lambda^2+2(1-\tau\tau')p_F^2\big]
\mathrm{ln}\big( 1+4p_F^2/\Lambda^2\big)\nonumber\\
&&\phantom{aaaaaa}+\Big( \frac{1}{2}\Lambda p_F^2+\frac{1}{24}\Lambda^3 \Big) 
\Big[ \frac{2\tau\tau'-6}{9}\frac{p_F^2}{\Lambda^2+4p_F^2 }\nonumber\\
&&\phantom{aaaaaa}-\frac{2}{9}(1-\tau\tau')\frac{p_F^2}{\Lambda^2}-\frac{16}{9}(1+\tau\tau')\frac{p_F^4}{(\Lambda^2+4p_F^2)^2}\Big]. \nonumber
\end{eqnarray}
Owing to the symmetry with respect to the interchange of isospin labels of the integral in \eqref{int2fermispheres}
it can easily be shown that there are no contributions to the equation of state from terms proportional to odd powers
of the isospin asymmetry $\beta$ provided that the interaction is charge symmetric ($C_{1,1}$=$C_{-1,-1}$)
since the following identities hold true
\begin{eqnarray}
 &&S^{(2n+1)}_{1,-1}\equiv 0, \\ &&S^{(2n+1)}_{-1,1}\equiv 0, \nonumber\\
&&S^{(2n+1)}_{1,1}+S^{(2n+1)}_{-1,-1}\equiv 0. \nonumber
\end{eqnarray}

%
% BibTeX users please use
\bibliography{references}
\bibliographystyle{epj}
%

%% Non-BibTeX users please use
%\begin{thebibliography}{}
%%
%% and use \bibitem to create references.
%Author, \textit{Book title} (Publisher, place year) page numbers
%\end{thebibliography}

\end{document}